\documentclass{aa} 
\usepackage{natbib}
\usepackage{graphicx} 
\usepackage{txfonts}
\usepackage{longtable} 
\usepackage{supertabular} 
\begin{document}

\newcommand{\rsc}{$r_{cs}$} 
\newcommand{\ntot}{$N_{tot}$}
\newcommand{\nbkg}{$N_{bkg}$} 
\newcommand{\nmem}{$N_{mem}$}
\newcommand{\nfield}{$N_{field}$}
\newcommand{\parchi}{$\chi^2_S$} 
\newcommand{\parchithresh}{$\chi^2_{S,threshold}$}
\newcommand{\be}{\begin{equation}} 
\newcommand{\ee}{\end{equation}}

\title{Substructures in WINGS clusters
\thanks{Figure 6 is only available in electronic form via 
http://www.edpsciences.org} } 
\author{M. Ramella\inst{1} \and A. Biviano\inst{1} \and A. Pisani\inst{2} \and
J. Varela\inst{3} \and D. Bettoni\inst{3} \and W.J.
Couch\inst{4} \and 
M. D'Onofrio\inst{5} \and A. Dressler\inst{6} 
\and G. Fasano\inst{3} \and P. Kj{\ae}rgaard\inst{7} \and M. Moles\inst{8} \and E. Pignatelli\inst{3} 
\and B.M. Poggianti\inst{3}
}

\offprints{Massimo Ramella, ramella@oats.inaf.it}

\institute{INAF/Osservatorio Astronomico di Trieste, via G. B. Tiepolo 11, I-34131, Trieste, Italy \and 
Istituto di Istruzione Statale Classico Dante Alighieri,
Scientifico Duca degli Abruzzi, Magistrale S. Slataper, 
viale XX settembre 11, I-34170 Gorizia, Italy \and
INAF/Osservatorio Astronomico di Padova, vicolo Osservatorio 5, I-35122, Padova, Italy \and 
School of Physics, University of New South Wales, Sydney 2052, Australia \and
Dipartimento di Astronomia, Universit\`a di Padova, vicolo Osservatorio 2, I-35122 Padova, Italy \and 
Observatories of the Carnegie Institution of Washington, Pasadena, CA 91101, USA \and
Copenhagen University Observatory. The Niels Bohr Institute for Astronomy Physics and Geophysics, Juliane Maries Vej 30, 2100 Copenhagen, Denmark \and
Instituto de Astrof\'{\i}sica de Andaluc\'{\i}a (C.S.I.C.) Apartado 3004, 18080 Granada, Spain
}

\date{Received / Accepted}

\abstract{} {We search for and characterize substructures in the
projected distribution of galaxies observed in the wide field CCD
images of the 77 nearby clusters of the WIde-field Nearby
Galaxy-cluster Survey (WINGS). This sample is complete
in X-ray flux in the redshift range $0.04<z<0.07$.}  {We search
for substructures in WINGS clusters with DEDICA, an adaptive-kernel
procedure. We test the procedure on Monte-Carlo simulations of the
observed frames and determine the reliability for the detected
structures.}  {DEDICA identifies at least one reliable structure in
the field of 55 clusters.  40 of these clusters have a total of 69
substructures at the same redshift of the cluster (redshift estimates
of substructures are from color-magnitude diagrams). The fraction of
clusters with subclusters (73\%) is higher than in most studies. 
The presence of subclusters affects the relative luminosities
of the brightest cluster galaxies (BCGs). Down
to $L \sim 10^{11.2} \, L_{\odot}$, our observed differential
distribution of subcluster luminosities is  consistent 
with the theoretical prediction of the differential mass function of
substructures in cosmological simulations.}{}


\titlerunning{Substructures in the WINGS clusters}
\authorrunning{M. Ramella et al.}

\maketitle

\section{Introduction}
According to the current cosmological paradigm, large structures in
the Universe form hierarchically. Clusters of galaxies are the largest
structures that have grown through mergers of smaller units and have
achieved near dynamical equilibrium. In the hierarchical scenario,
clusters are a rather young population, and we should be able to
observe their formation process even at rather low redshifts. A
signature of such process is the presence of cluster substructures. A
cluster is said to contain substructures (or subclusters) when its
surface density is characterized by multiple, statistically
significant peaks on scales larger than the typical galaxy size, with
``surface density'' being referred to the cluster galaxies, the
intra-cluster (IC) gas or the dark matter \citep[DM
hereafter;][]{Buote02}.

Studying cluster substructure therefore allows us to investigate the
process by which clusters form, constrain the cosmological model of
structure formation, and ultimately test the hierarchical paradigm
itself \citep[e.g.][]{RLT92,MEFG95,Thomas+98}. In addition, it also
allows us to better understand the mechanisms affecting galaxy
evolution in clusters, which can be accelerated by the perturbative
effects of a cluster-subcluster collision and of the tidal field
experienced by a group accreting onto a cluster
\citep{Bekki99,Dubinski99,Gnedin99}. If clusters are to be used as
cosmological tools, it is important to calibrate the effects
substructures have on the estimate of their internal properties
\citep[e.g.][]{SM93,Pinkney+96,RSM98,Biviano+06,Lopes+06}. Finally,
detailed analyses of cluster substructures can be used to constrain
the nature of DM \citep{Markevitch+04,Clowe+06}.

The analysis of cluster substructures can be performed using the
projected phase-space distribution of cluster galaxies
\citep[e.g.][]{GB82}, the surface-brightness distribution and
temperature of the X-ray emitting IC gas \citep[e.g.][]{BHB92}, or the
shear pattern in the background galaxy distribution induced by
gravitational lensing, that directly samples substructure in the DM
component \citep[e.g.][]{ASW98}. None of these tracers of cluster
substructure (cluster galaxies, IC gas, background galaxies) can be
considered optimal. The identification of substructures is in fact
subject to different biases depending on the tracer used. In X-rays
projection effects are less important than in the optical, but the
identification of substructures is more subject to a $z$-dependent
bias, arising from the point spread function of the X-ray telescope
and detector \citep[e.g.][]{BS02}. Moreover, the different cluster
components respond in a different way to a cluster-subcluster
collision. The subcluster IC gas can be ram-pressure braked and
stripped from the colliding subcluster and lags behind the subcluster
galaxies and DM along the direction of collision
\citep[e.g.][]{RLB97,Barrena+02,Clowe+06}. Hence, it is equally useful
to address cluster substructure analysis in the X-ray and in the
optical.

Traditionally, the first detections of cluster substructures were
obtained from the projected spatial distributions of galaxies
\citep[e.g.][]{SW54,ANS64}, in combination, when possible, with the
distribution of galaxy velocities
\citep[e.g.][]{vandenBergh60,vandenBergh61,deVaucouleurs61}. 
Increasingly sophisticated techniques for the detection and
characterization of cluster substructures have been developed over the
years \citep[see][and references
therein]{Moles+86,Perea+86a,Perea+86b,Buote02,GB02}.
In many of these
techniques substructures are identified as deviations from symmetry in
the spatial and/or velocity distribution of galaxies and in the X-ray
surface-brightness \citep[e.g.][]{WOD88,FM88,MFG93,SBRF01}. In other
techniques substructures are identified as significant peaks in the
surface density distribution of galaxies or in the X-ray surface
brightness, either as residuals left after the subtraction of a
smooth, regular model representation of the cluster
\citep[e.g.][]{NB97,EFW98}, or in a non-parametric way, e.g. by the
technique of wavelets \citep[e.g.][]{Escalera+94,SDG94,Biviano+96} and
by adaptive-kernel techniques
\citep[e.g.][]{KB97,Bardelli+98a,BZB01}.

The performances of several different methods have been evaluated both
using numerical simulations
\citep[e.g.][]{MEFG95,CER96,Pinkney+96,BX97,Cen97,VGB99,KM00,Biviano+06}
and also by applying different methods to the same cluster data-sets
and examine the result differences
\citep[e.g.][]{ESM92,Escalera+94,MEFG95,Mohr+96,KB97,FSB98,KBPG01,SBRF01,Lopes+06}.
Generally speaking, the sensitivity of substructure detection
increases with both increasing statistics (e.g. more galaxies or more
X-ray photons) and increasing dimensionality of the test (e.g. using
galaxy velocities in addition to their positions, or using X-ray
temperature in addition to X-ray surface brightness).

Previous investigations have found very different fractions of clusters with
substructure in nearby clusters, depending on the method and tracer 
used for substructure detection, on the cluster sample, and 
on the size of sampled cluster
regions
\citep[e.g.][]{GB82,DS88,MEFG95,Girardi+97,KB97,JF99,SSG99,KBPG01,SBRF01,FK06,Lopes+06}. Although
the distribution of subcluster masses has not been determined
observationally, it is known that subclusters of $\sim 10$\% the
cluster mass are typical, while more massive subclusters are less
frequent \citep{Escalera+94,Girardi+97,JF99}. The situation is
probably different for distant clusters which tend to show massive
substructures more often than nearby clusters clearly suggesting
hierarchical growth of clusters was more intense in the past
\citep[e.g.][]{Gioia+99,vanDokkum+00,HCCA01,Maughan+03,Huo+04,Rosati+04,Demarco+05,JCBB05}. 

Additional evidence for the hierarchical formation of clusters is
provided by the analysis of brightest cluster galaxies (BCGs hereafter)
in substructured clusters. BCGs usually sit at the bottom of the
potential well of their host cluster \citep[e.g.][]{Adami+98a}. When a
BCG is found to be significantly displaced from its cluster dynamical
center, the cluster displays evidence of substructure
\citep[e.g.][]{Beers+91,Ferrari+06}. From the correlation between
cluster and BCG luminosities, \citet{LM04} conclude that BCGs grow by
merging as their host clusters grow hierarchically. The related
evolution of BCGs and their host clusters is also suggested by the
alignement of the main cluster and BCG axes
\citep[e.g.][]{Binggeli82,Durret+98}. Both the BCG and the cluster
axes are aligned with the surrounding large scale structure
distribution, where infalling groups come from. These infalling groups
are finally identified as substructures once they enter the cluster
environment
\citep{Durret+98,Arnaud+00,WB00,Ferrari+03,Plionis+03,Adami+05}. Hence,
substructure studies really provide direct evidence for the
hierarchical formation of clusters.

Concerning the impact of subclustering on global cluster properties,
it has been found that subclustering leads to over-estimating cluster
velocity dispersions and virial masses
\citep[e.g.][]{Perea+90,Bird95,Maurogordato+00},
but not in the general case of
small substructures \citep{Escalera+94,Girardi+97,XFW00}. During the
collision of a subcluster with the main cluster, both the X-ray
emitting gas distribution and its temperature have been found to be
significantly affected \citep[e.g.][]{MV01,Clowe+06}. As a
consequence, it has been argued that substructure can explain at least
part of the scatter in the scaling relations of optical-to-X-ray
cluster properties
\citep[e.g.][]{Fitchett88,Girardi+96,Barrena+02,Lopes+06}.

As far as the internal properties of cluster galaxies are concerned,
there is observational evidence that a higher fraction of cluster
galaxies with spectral features characteristic of recent or ongoing
starburst episodes is located in substructures or in the regions of
cluster-subcluster interactions
\citep{Caldwell+93,Abraham+96,Biviano+97,CR97,Bardelli+98b,MW00,Miller+04,Poggianti+04,Miller05,Giacintucci+06}.

In this paper we search for and characterize substructures in the
sample of 77 nearby clusters of the WIde-field Nearby Galaxy-cluster
Survey \citep[WINGS hereafter,][]{Fasano+06}. This sample is an almost
complete sample in X-ray flux in the redshift range $0.04<z<0.07$.
We detect substructures from the spatial, projected distribution of 
galaxies in the cluster fields, using the adaptive-kernel based
DEDICA algorithm \citep{Pisani93,Pisani96}. In Sect.~\ref{s:data}
we describe our data-set; in Sect.~\ref{s:dedica} we describe the
procedure of substructure identification; in Sect.~\ref{s:sims} we
use Monte Carlo simulations in order to tweak our procedure; in
Sect.~\ref{s:detection} we describe the identification of substructures in our
data-set; in Sect.~\ref{s:catalog} the catalog of identified substructures is
provided. In Sect.~\ref{s:properties} we investigate the properties of the
identified substructures, and in Sect.~\ref{s:bcg} we consider the relation
between the BCGs and the substructures. We provide a summary of our work in
Sect.~\ref{s:summary}. 

\section{The Data}
\label{s:data}

WINGS is an
all-sky, photometric (multi-band) and spectroscopic survey, whose
global goal is the systematic study of the local cosmic variance of
the cluster population and of the properties of cluster galaxies as a
function of cluster properties and local environment.

The WINGS sample consists of 77 clusters selected from three X-ray
flux limited samples compiled from ROSAT All-Sky Survey data, with
constraints just on the redshift ($0.04< z<0.07$) and distance from
the galactic plane (${\vert}b{\vert}\geq$20~deg).
The core of the project consists of wide-field optical imaging of
the selected clusters in the $B$ and $V$ bands. The imaging data were 
collected using the WFC@INT (La Palma) and the WFI@MPG (La Silla) 
in the northern and southern hemispheres, respectively.

The observation strategy of the survey favors the uniformity 
of photometric depth inside the different CCDs, rather than complete
coverage of the fields that would require dithering. Thus, the 
gaps in the WINGS optical imaging correspond to the  physical 
gaps between the different CCDs of the mosaics.

During the data reduction process, we give particular care 
to sky subtraction (also in presence of
crowded fields including big halo galaxies and/or very bright stars), 
image cleaning (spikes and bad pixels) and 
star/galaxy classification (obtained with both automatic and 
interactive tools).

According to \citet{Fasano+06} and \citet{Varela+07}, the
overall quality of the data reported in the WINGS photometric catalogs
can be summarized as follows: (i) the astrometric errors for extended
objects have $r.m.s.\sim$0.2 arcsec; (ii) the average limiting magnitude is
$\sim$24.0, ranging from 23.0 to 25.0; (iii) the completeness of the
catalogs is achieved (on average) up to $V\sim$22.0; (iv) the total
(systematic plus random) photometric $r.m.s.$ errors, derived from
both internal and external comparisons, vary from $\sim$0.02~mag, for
bright objects, up to $\sim$0.2~mag, for objects close to the
detection limit.

\section{The DEDICA Procedure}
\label{s:dedica}

We base our search for substructures in WINGS clusters on the DEDICA
procedure \citep{Pisani93,Pisani96}. This procedure has the following
advantages:
\begin{enumerate}
\item DEDICA gives a total description of the clustering pattern, in
particular the membership probability and significance of structures besides
geometrical properties;
\item DEDICA is scale invariant;
\item DEDICA does not assume any property of the clusters, i.e. it
is completely non-parametric. In particular it does not require
particularly rich samples to run effectively.
\end{enumerate}
The basic nature and properties of DEDICA are described in
\citet[][and references therein]{Pisani93,Pisani96}.  Here we
summarize the main structure of the algorithm and how we apply it to
our data sample. The core structure of DEDICA is
based on the assumption that a structure (or a ``cluster'' in the 
algorithm jargon) corresponds to a local maximum
in the density of galaxies.

We proceed as follows.  First we need to estimate the
probability density function $\Psi({\bf r}_{i})$ (with $i=1,\ldots N$)
associated with the set of N galaxies with coordinates ${\bf r}_{i}$. Second,
we need to find the local maxima in our estimate of $\Psi({\bf r}_{i})$ in
order to identify clusters and also to evaluate their significance relatively
to the noise.  Third and finally, we need to estimate the probability that a
galaxy is a member of the identified clusters.

\subsection{The probability density}
\label{ss:probdens}

DEDICA is a non-parametric method in the sense that it does not require any
assumption on the probability density function that it is aimed to estimate.
The only assumptions are that  $\Psi({\bf r}_{i})$ must be continuous and at
least twice differentiable.

The function $f({\bf r}_{i})$ is an estimate of $\Psi({\bf r}_{i})$ and it is
built by using an adaptive kernel method given by: 

\be f_{ka}({\bf r})=\frac{1}{N} \sum_{i=1}^{N} K({\bf r}_{i},\sigma_{i};{\bf
r}) \label{f_ka} \ee where we use the two dimensional Gaussian kernel $K({\bf
r}_{i},\sigma_{i};{\bf r})$ centered in ${\bf r}_{i}$ with size $\sigma_{i}$.

The most valuable feature of DEDICA is the procedure to select the values of
kernel widths $\sigma_{i}$. It is possible to show that the optimal choice for
$\sigma_{i}$, i.e. with asymptotically minimum variance and 
null bias, is obtained by minimizing the distance between our estimate
$f({\bf r}_{i})$ and $\Psi({\bf r}_{i})$.  This distance can be evaluated by a
particular function called  the integrated square error $ISE(f)$ given by: \be
ISE(f)=\int_{\Re} [\Psi({\bf r}) - f({\bf r})]^2 d{\bf r} \ee.

Once the minimum $ISE(f)$ is reached we have obtained the DEDICA estimate of
the density as in Eq.\ref{f_ka}. 

\subsection{Cluster Identification}
\label{ss:clid}

The second step of DEDICA consists in the identification of the local
maxima in $f_{ka}({\bf r})$. The positions of the peaks in the density
function $f_{ka}({\bf r})$ are found as the solutions of the iterative
equation: \be {\bf r}_{m+1}={\bf r}_{m} +a \cdot \frac{\nabla
f_{ka}({\bf r}_m)}{f_{ka}({\bf r}_m)} \label{peak} \ee where $a$ is a
scale factor set according to optimal convergence requirements. The
limit ${\bf R}$ of the sequence ${\bf r}_m$ defined in Eq.\ref{peak}
depends on the starting position ${\bf r}_{m=1}$.  \be \lim_{m
\rightarrow + \infty} {\bf r}_{m}= {\bf {\bf R}}( {\bf r}_{m=1}) \ee
We run the sequence in Eq.\ref{peak} at each data position ${\bf
r}_i$. We label each data point with the limit ${\bf {\bf R}}_i={\bf
{\bf R}}({\bf r}_{m=1}={\bf r}_{i})$. These limits ${\bf R}_i$ are the
position of the peak to which the $i-th$ galaxy belong.  In the case
that all the galaxies are members of a unique cluster, all the labels
${\bf R}_i$ are the same. At the other extreme each galaxy is a
one-member cluster and all ${\bf R}_i$ have different values. All the
members of a given cluster belong to the same peak in $f_{ka}({\bf
r})$ and have the same ${\bf R}_i$. We identify cluster members by
listing galaxies having the same values of ${\bf R}$. We end up with
$\nu$ different clusters each with $n_{\mu}$ ($\mu=1, \ldots, \nu$)
members.
 
In order to maintain a coherent notation, we identify with the label $\mu=0$
the $n_0$ isolated galaxies considered a system of background galaxies. We
have: $n_0 = N- \sum_{\mu=1}^{\nu} n_\mu$.

\subsection{Cluster Significance and of Membership Probability}
\label{ss:membprob}
The statistical significance $S_\mu$ ($\mu=1,\ldots,\nu$) of each cluster is
based on the assumption that the presence of the $\mu-th$ cluster causes an
increase in the local probability  density as well as  in the sample likelihood
$L_N = \Pi_i [ f_{ka}({\bf r}_i) ]$ relatively to the value $L_\mu$ that one
would have if the members of the $\mu-th$ cluster were all isolated, i.e.
belonging to the background. 

A large value in the ratio $L_N / L_\mu$ characterizes the most
important clusters. According to \citet{Materne79} it is possible to
estimate the significance of each cluster by using the likelihood
ratio test. In other words $2 \ln (L_N / L_\mu)$ is distributed as a
$\chi^2$ variable with $\nu-1$ degrees of freedom. Therefore, once we
compute the value of $\chi^2$ for each cluster (\parchi), we can also
compute the significance $S_\mu$ of the cluster.

Here we assume that the contribution to the global density field $f_{ka}({\bf
r}_i)$ of the $\mu-th$ cluster is $F_\mu({\bf r}_i)$.  The
ratio between the value of $F_\mu({\bf r}_i)$ and the total local density
$f_{ka}({\bf r}_i)$ can be used to estimate the membership probability of each
galaxy relatively to the identified clusters.  This criterion also allows us to
estimate the probability that a galaxy is isolated.

At the end of the DEDICA procedure we are left with a) a catalog of
galaxies each with information on position, membership, local density
and size of the Gaussian kernel, b) a catalog of structures with
information on position, richness, the \parchi parameter, and peak
density.  For each cluster we also compute from the coordinate
variance matrix the cluster major axis, ellipticity and position
angle.

\section{Tweaking the Algorithm with Simulations}
\label{s:sims}

In this section we describe our analysis of the performance of DEDICA and the
guidelines we obtain for the interpretation of the clustering analysis of our
observations.

We build simulated fields containing a cluster with and without
subclusters. The simulated fields have the same geometry of the WFC
field and are populated with the typical number of objects we will
analyze.  For simplicity we consider only WFC fields. Because DEDICA
is scale-free, a different sampling of the same field of view has no
consequence on our analysis.

In the next section we limit our analysis to $M_{V,lim} \leq -16$.
At the median redshift of the WINGS cluster, $z \simeq 0.05$, this
absolute magnitude limit corresponds to an apparent
 magnitude  $V_{lim} \simeq  21$.
Within this magnitude limit the representative number of galaxies in
our frames is \ntot = 900.

We then consider \ntot = \nmem + \nbkg, with \nmem the number of
cluster members and \nbkg the number of field -- or background --
galaxies.  We set \nbkg = 670, close to the average number of
background galaxies we expect in our frame based on typical observed
fields counts, e.g. \citet{Berta+06} or \citet{Arnouts+97}.  With this
choice, we have \nmem = 230.

We distribute uniformly at random \nbkg objects. We distribute at
random the remaining \nmem = 230 objects in one or more overdensities
depending on the test we perform. We populate overdensities according
to a King profile \citep{King62} with a core radius $R_{core} = 90$ kpc,
representative of our clusters.  We then scale $R_{core}$ with the
number of members of the substructure, $N_S$. We use

$$R_{core} = 250 \sqrt{ \frac{N_S}{N_C + N_S}}$$ 

where $N_C$ is the number of objects in the cluster with \nmem = $N_S
+ N_C$.  This scaling of $R_{core}$ with cluster richness is from
\citet{Adami+98b} assuming direct proportionality between cluster
richness and luminosity \citep[e.g.][]{PBBR06}.

As far as the relative richnesses of the cluster and subcluster are
concerned, we consider the following richness ratios $r_{cs} =
N_{C}/N_{S} = 1, 2, 4, 8$. With these richness ratios, 
the number of objects in the
cluster are $N_{C} = 115, 153, 184, 204$, and those in
subclusters are $N_{S} = 115, 77, 46, 26$ respectively.

In a first set of simulated fields we place the substructure at 2731 pixels (15
arcmin) from the main cluster so that they do not overlap.  In a second set of
simulations, we place main cluster and substructure at shorter distances, 683
and 1366 pixels, in order to investigate the ability of DEDICA to resolve
structures.  At each of these shorter distances we build simulations 
with both \rsc = 1 and 2. 

For each richness ratio and/or distance between cluster and subcluster we
produce 16 simulations with different realizations of the random positions of
the data points representing galaxies.

In order to minimize the effect of the borders on the detection of structures
we add to the simulation a ``frame'' of 1000 pixel.  We fill this frame with a
grid of data points at the same density as the average density of the field.

The first result we obtain from the runs of DEDICA on the simulations with
varying richness ratio is the positive rate at which we detect real structures.
We find that we always recover both cluster and substructure even when the
substructure only contains $N_S$ = 1/8 $N_C$ objects, i.e. 26 objects (on top
of the uniform background). In other words, if there is a real structure DEDICA
finds it.

\begin{figure} \centering \includegraphics[width=8.8cm]{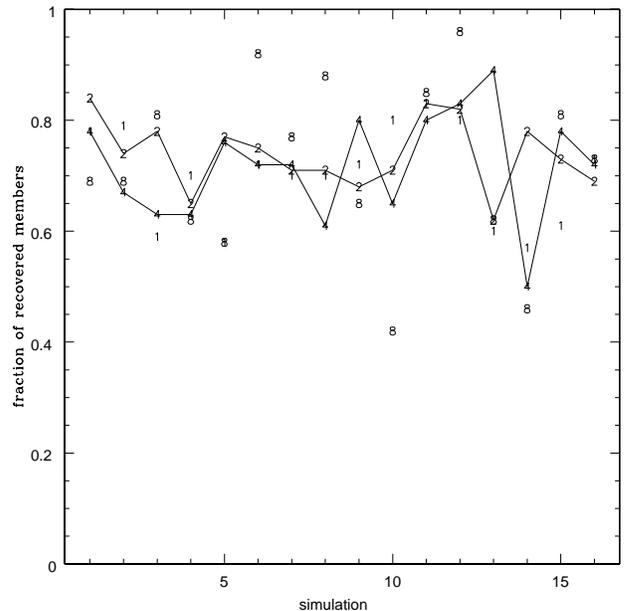}
\caption{Fraction of recovered members of each substructure for different
\rsc. The solid line connects substructures with $r_{cs}$ = 2 and 4} 
\label{fig:recovery} \end{figure}

We also check how many original members the procedure assigns to structures it
recovers. The results are summarized in Fig.~\ref{fig:recovery}. In the
diagram, the fraction of recovered members of each substructure is represented
by the values of its $r_{cs}$.  The solid line connects substructures with
$r_{cs}=2$ and 4. 

From Fig.~\ref{fig:recovery} it is clear that our procedure recovers a large
fraction  of members, almost irrespective of the richness of the original
structure.  It is also interesting to note that the fluctuations identified as
substructures are located very close to the center of the corresponding
simulated substructures. In almost all cases the distance between original and
detected substructure is significantly shorter than the mean inter-particle
distance.

The second important result we obtain from the simulations is the false
positive rate, i.e. the fraction of noise fluctuations that are as significant
as the fluctuations corresponding to real structures. 

First of all we need to define an operative measure of the
reliability of the detected structures. In fact DEDICA provides a
default value $S_\mu$ ($\mu=1,\ldots,\nu$) of the significance (see
Sect.~\ref{ss:membprob}). However, $S_\mu$ has a
relatively small dynamical range, in particular for highly significant
clusters.

Density or richness both allow a reasonable "ranking" of structures.  However,
both large low-density noise fluctuations (often built up from more than one
noise fluctuation) and very high density fluctuations produced by few very
close data points could be mistakenly ranked as highly significant structures
according to, respectively, richness and density criteria. 

We therefore prefer to use the parameter \parchi which stands at the base of
the estimate of $S_\mu$ and which is naturally provided by DEDICA. The main
characteristic of \parchi is that it depends {\it both} on the density of a
cluster relative to the background and on its richness. Using  \parchi we
classify correctly significantly more structures than with either density or
richness alone.
 
\begin{figure} \centering \includegraphics[width=8.8cm]{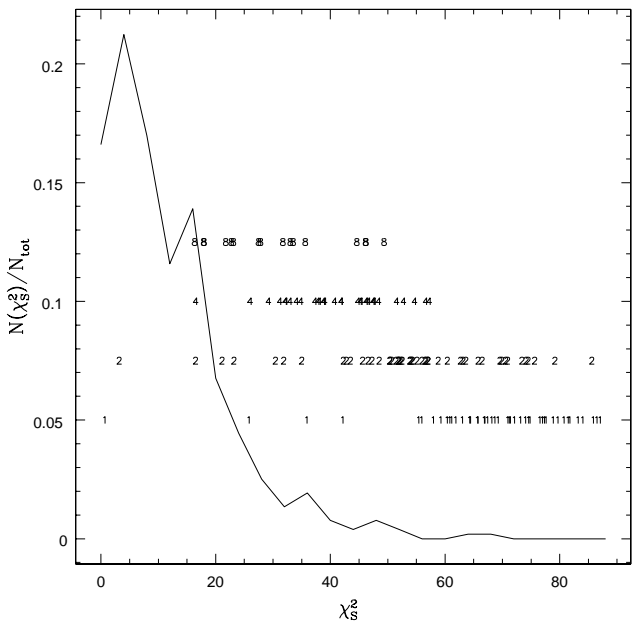}
\caption{\parchi of simulated noise fluctuations (solid line). Labels
are the \rsc of simulated structures at the abscissa corresponding to their
\parchi and at arbitrary ordinates.} \label{fig:flucthisto} \end{figure}

In Fig.~\ref{fig:flucthisto} we plot the distribution of \parchi of noise
fluctuations (solid line). In the same plot we also mark the \rsc of real
structures as detected by our procedure. We use labels indicating \rsc and
place them at the abscissa corresponding to their
\parchi and at arbitrary ordinates.

Fig.~\ref{fig:flucthisto} shows that the structures detected with \rsc
= 1, 2 are always distinguishable from noise fluctuations.
Substructures with \rsc = 4 or higher, although correctly detected,
have \parchi values that are close to or lower than the level of
noise.

With the second set of simulations, we test the minimum distance at
which cluster and subcluster can still be identified as separate
entities.  We place cluster and substructure (\rsc = 1, 2) at
distances $d_{cs} = $ 683 and 1366 pixel. These distances are 1/4 and
1/2 respectively of the distance between cluster and substructure in
the first set of simulations. Again we produce 16 simulations for each
of the 4 cases.

We find that at $d_{cs} = $ 1366 pixel cluster and substructure are always
correctly identified. At the shorter distance $d_{cs} = $ 683 pixel, DEDICA
merges cluster and substructure in 1 out of 16 cases for \rsc = 1 and in 8 out
of 16 cases for \rsc = 2.  With our density profile, $d_{cs} = $ 683 pixel
corresponds to $d_{cs} \simeq R_c + R_s$ with $ R_c$, $ R_s$ the radii 
of the main cluster and of the subcluster respectively.

In order to verify the results we obtain for 900 data points we produce more
simulations with \ntot = 450, 600 and 1200. In all these simulations $R_C$ and
$R_S$ are the same as in the set with \ntot = 900.  We vary \nbkg and \nmem so
that \nmem / \nbkg is the same as in the case \ntot = 900. 

These simulations  confirm the results we obtain in the case
\ntot = 900, and allow us to set a detection
threshold, \parchithresh(\ntot), for significant fluctuations 
in the analysis of real clusters. 

\begin{figure} \centering \includegraphics[width=8.8cm]{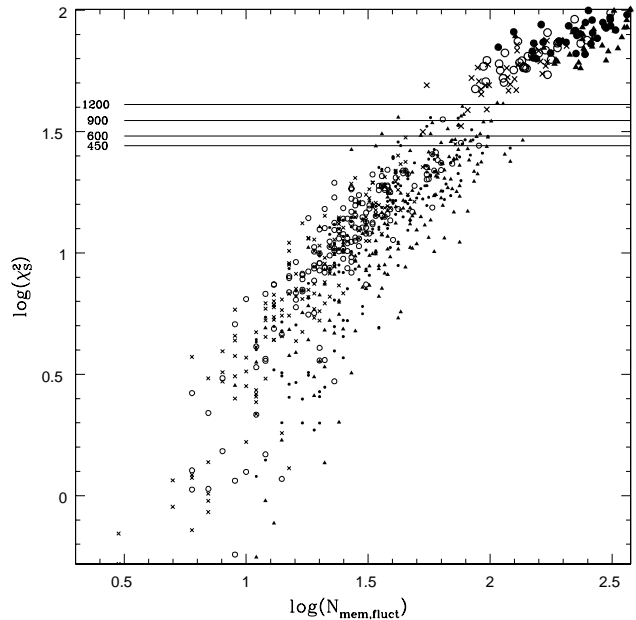}
\caption{Small symbols correspond to \parchi as a function of the
number of members of noise fluctuations. Crosses, circles, dots and
triangles are \parchi for the noise fluctuations of the simulations
with \ntot = 450, 600, 900, and 1200 respectively. Large symbols are
\parchi of simulated clusters and subclusters with \rsc = 1.
Horizontal lines mark the levels of \parchithresh.}
\label{fig:parchinoise} \end{figure}

We summarize the behavior of the noise fluctuations in our simulations
in Fig.~\ref{fig:parchinoise}. In this figure, the small symbols
correspond to \parchi as a function of the number of members of noise
fluctuations. In particular, crosses, circles, dots and triangles are
\parchi for the noise fluctuations of the simulations with \ntot =
450, 600, 900, and 1200 respectively.

The larger symbols are the \parchi of the fluctuations corresponding to
simulated clusters and subclusters of equal richness (\rsc = 1).

The 4 horizontal lines mark the level of \parchithresh, i.e. the
average \parchi of the 3 most significant noise fluctuations in each
of the 4 groups of simulations with \ntot = 450, 600, 900, and 1200.

The expected increase of \parchithresh with \ntot is evident. 

We note that the only significant difference with these findings we obtain from
the simulations with \rsc = 2 is that \parchi of simulated clusters and
subclusters is closer to \parchithresh (but still higher).

We fit \parchithresh with \ntot and obtain 

\begin{equation} \log(\chi^2_{S,threshold}) = 1/2.55 ~ \log(N_{tot}) + 0.394
\label{eq:noisethresh} \end{equation} in good agreement with the expected
behavior of the poissonian fluctuations.  

As a final test we verify that infra-chip gaps do not have a dramatic impact on
the detection of structures in the cases \rsc = 1 and 2. We place a 50
pixel wide gap where it has the maximum impact, i.e. where the
kernel size is shortest.  Even if the infra-chip gap cuts through the center of
the structures, DEDICA is able to identify these structures correctly. 

We summarize here the main results of our tests on simulated
clusters with substructures:

\begin{itemize}
\item  DEDICA successfully detects even the poorest structures above a uniform 
poissonian noise background.
\item  DEDICA recovers a large fraction (typically $>$ 3/4) of the real members 
of a substructure, almost irrespective of the richness of the structure.
\item  DEDICA is able to distinguish between noise fluctuations and  
true structures only if these structures are rich enough. In the case of our
simulations, structures have to be richer than 1/4th of the main structure.
\item  DEDICA is able to separate neighboring structures provided they 
do not overlap.  
\item  infra-chip gaps do not threaten the detection of structures that
are rich enough to be reliably detected.
\item  the \parchi threshold we use to identify significant 
structures is a function of the total number of points and can be 
scaled within the whole range of numbers of galaxies observed
within our fields.  
\end{itemize}

In the next section we apply these results to the real WINGS clusters.

\section{Substructure detection in WINGS clusters}
\label{s:detection}

We apply our clustering procedure to the 77 clusters of the WINGS sample. The
photometric catalog of each cluster is deep, reaching a completeness
magnitude $V_{complete} \leq 22$. The number of galaxies is correspondingly
large, from $N_{gal} \simeq 3,000 $ to $N_{gal} \simeq 10,000$. 

The large number of bright background galaxies (faint apparent
magnitudes) dilutes the clustering signal of local WINGS clusters.  We
perform test runs of the procedure on several clusters with magnitude
cuts brighter than $V_{complete}$.  Based on these tests, we decide to
cut galaxy catalogs to the absolute magnitude threshold $M_V =
-16.0$. With this choice a) we maximize the signal-to-noise ratio of
the detected subclusters and b) we still have enough galaxies for a
stable identification of the system.  At the median redshift of WINGS
clusters, $z \simeq 0.0535$, our absolute magnitude cut corresponds to
an apparent magnitude $V \simeq 21.2$.

This apparent magnitude also approximately corresponds to the magnitude
where the contrast of our typical cluster relative to the field is maximum
(this estimate is based on the average cluster luminosity function of
\citep{Yagi+02,DePropris+03} and on the galaxy counts of \citep{Berta+06}).

The number of galaxies that are brighter than the threshold $M_V =
-16.0$ is in the range $600 < N_{tot} < 1200$ for a large fraction of
clusters observed with either WFC@INT or with WFI@ESO2.2.

In order to proceed with the identification of significant structures within
WINGS clusters, we need to verify that our simulations are sufficiently
representative of the real cases. In practice we need to compare the observed
distributions of \parchi values of noise fluctuations with the corresponding
simulated distributions. In the observations it is impossible to identify
individual fluctuations as noise. In order to have an idea of the distributions
of \parchi of noise fluctuations we consider that our fields are centered
on real clusters. As a consequence, on average, fluctuations in the center of
the frames are more likely to correspond to real systems than those at the
borders. 

\begin{figure} \centering \includegraphics[width=8.8cm]{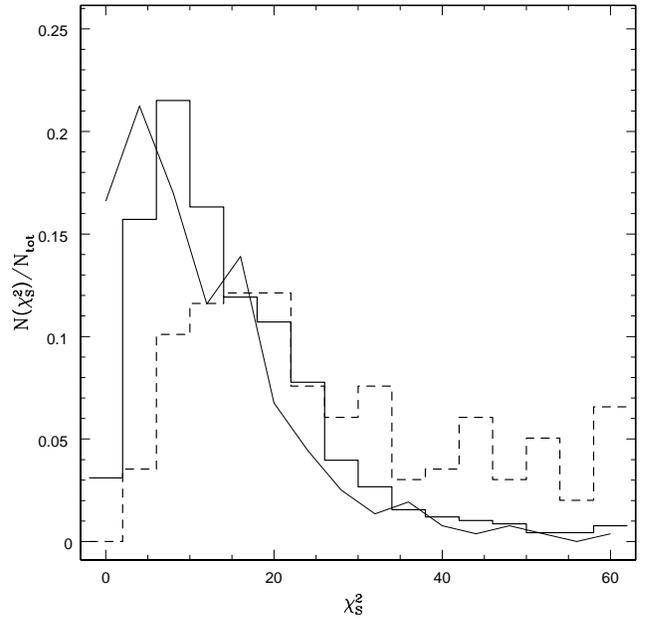}
\caption{\parchi distributions for border (thick solid histogram) and central
(thick dashed histogram) observed fluctuations.  The thin solid line is the
normalized distribution of \parchi of the noise fluctuations in our
simulations} \label{fig:chi2_border} \end{figure}

We therefore consider separately the fluctuations within the central
regions of the frames and all other fluctuations (borders).  We define
the central regions as the central 10\% of WFC and WFI areas.  We
plot in Fig.~\ref{fig:chi2_border} the two distributions. The thick
solid histogram is for the border and the thick dashed histogram for
the center of the frames.  The difference between "noise" and "signal"
is clear.  In the same figure we also plot the normalized distribution
of \parchi of the noise fluctuations in our simulations (thin solid
line). The distributions of \parchi of the observed and simulated
fluctuations are in reasonable agreement considering a) the simple
model used for the simulations and that b) in the observations we can
not exclude real low-\parchi structures among noise fluctuations. We
conclude that for our clusters we can adopt the same reliability
threshold \parchithresh we determine from our simulations
(Eq.~\ref{eq:noisethresh}).

\section{The Catalog of Substructures}
\label{s:catalog}

We detect at least one significant structure in 55 (71\%) clusters. We find
that 12 clusters (16\%) have no structure above the threshold (undetected).  In
the case of another 10 (13\%) clusters we find significant 
structures only at the
border of the field of view. In absence of a detection in the center of the
frame, we consider these border structures unrelated to the target cluster.  We
also verify that in the Color-Magnitude Diagram (CMD) these border
structures are redder than expected given the redshift of the target cluster. 
We consider also these 10 clusters undetected.

Here we list the 22 undetected clusters: A0133, A0548b, A0780, A1644, A1668,
A1983, A2271, A2382, A2589, A2626, A2717, A3164, A3395, A3490, A3497, A3528a,
A3556, A3560, A3809, A4059, RX1022, Z1261.

We note that undetected clusters are real physical systems according to their
x-ray selection. From an operative point of view, the fact that these clusters are 
not detected by DEDICA is the result of the division into too many structures 
of  the total available clustering signal in the field (or of a too large 
fraction of the clustering signal going into border structures). Several
physical situations could be at the origin of missed detections. One possibility
is an excess of physical substructures of comparable richness. Another possibility
is that these clusters are embedded in regions of the large scale structure that are highly
clustered.     

We do not try to recover these structures because
they can not be prominent enough. Since our analysis is bidimensional, we
can only detect and use confidently the most prominent structures. Redshifts
are needed for a more detailed analysis of cluster substructures.

We list the 55 clusters with significant structures 
in Table \ref{tab:dedica}.
We give, for each substructure: (1) the name of the parent cluster; (2) the
classification of the structure as main (M), subcluster (S), or
background (B) together with their order number; (3) right ascension
(J2000), and (4) declination (J2000) in decimal degrees of the DEDICA
peak; the parameters of the ellipse we obtain from the variance matrix
of the coordinates of galaxies in the substructure, i.e.  (5) major
axis in arcminutes, (6) ellipticity, and (7) position angle in
degrees; (8) luminosity (see the next section); (9) \parchi.

\begin{figure}
\centering \resizebox{\hsize}{!}{\includegraphics{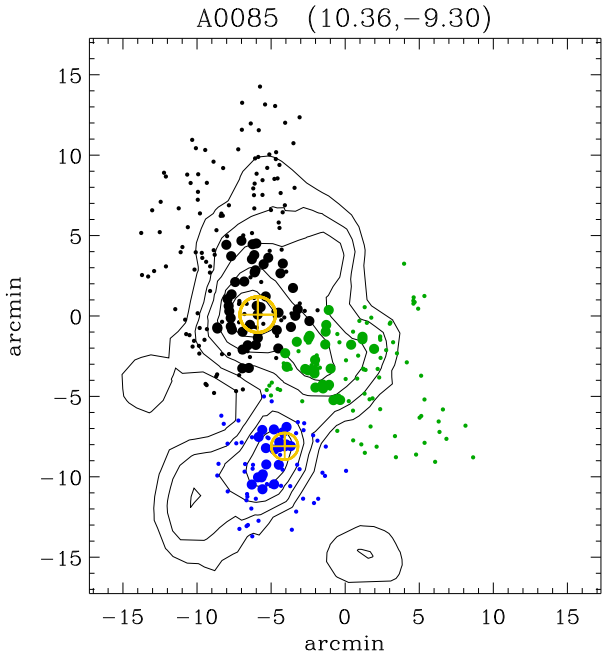}}
\caption{Isodensity contours (logarithmically spaced) of the Abell 85
field. The title lists the coordinates of the center. 
The orientation is East to the left, North to
the top. Galaxies belonging to 
the systems detected by DEDICA are shown
as dots of different colors. Black, light green, blue, red, magenta, 
dark green are for the main system and the subsequent 
substructures ordered as in
Table \ref{tab:dedica}. Large  symbols are for
galaxies with $M_V \leq -17.0$ that lie where local densities are higher than
the median local density of the structure the galaxy belongs to.
Open symbols mark the positions
of the first- and second-ranked cluster galaxies, BCG1 and BCG2
respectively. Similar plots for the 55 analysed clusters are
available in the electronic version of this Journal.}
\label{fig:a0085}
\end{figure}

\onlfig{6}{
\begin{figure*}
\centering 
\begin{minipage}{0.9\textwidth}
\resizebox{\hsize}{!}{\includegraphics{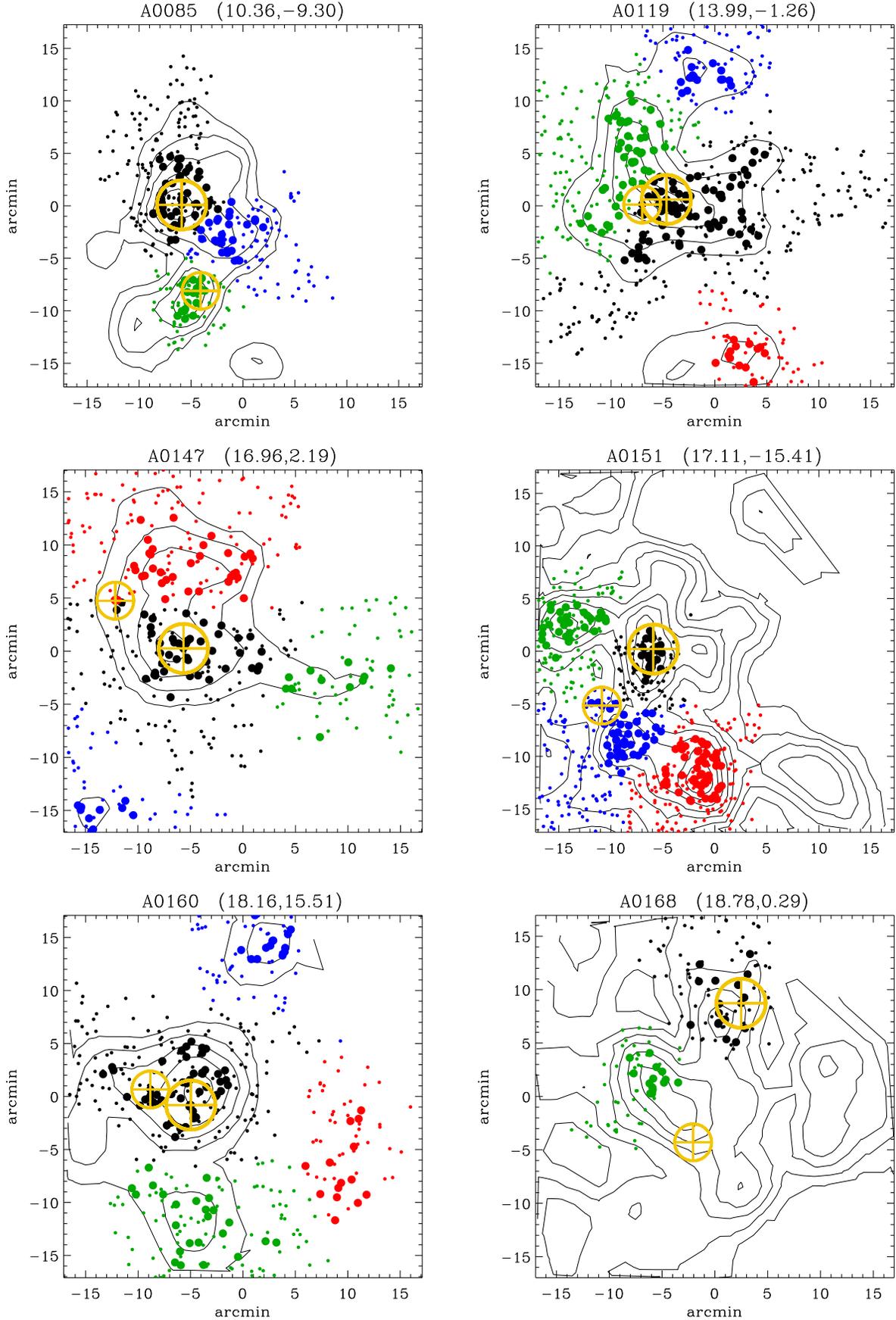}}
\end{minipage}
\caption{Isodensity contours (logarithmically spaced) of the 55
clusters with significant structures.
The title lists the coordinates of the center. 
The orientation is East to the left, North to
the top. Galaxies belonging to 
the systems detected by DEDICA are shown
as dots of different colors. Black, light green, blue, red, magenta, 
dark green are for the main system and the subsequent 
substructures ordered as in
Table \ref{tab:dedica}. Large  symbols are for
galaxies with $M_V \leq -17.0$ that lie where local densities are higher than
the median local density of the structure the galaxy belongs to.
Open symbols mark the positions
of the first- and second-ranked cluster galaxies, BCG1 and BCG2
respectively. 
}
\label{fig:online} 
\end{figure*}

\addtocounter{figure}{-1} 
\begin{figure*}  \centering
\begin{minipage}{0.9\textwidth}
\resizebox{\hsize}{!}{\includegraphics{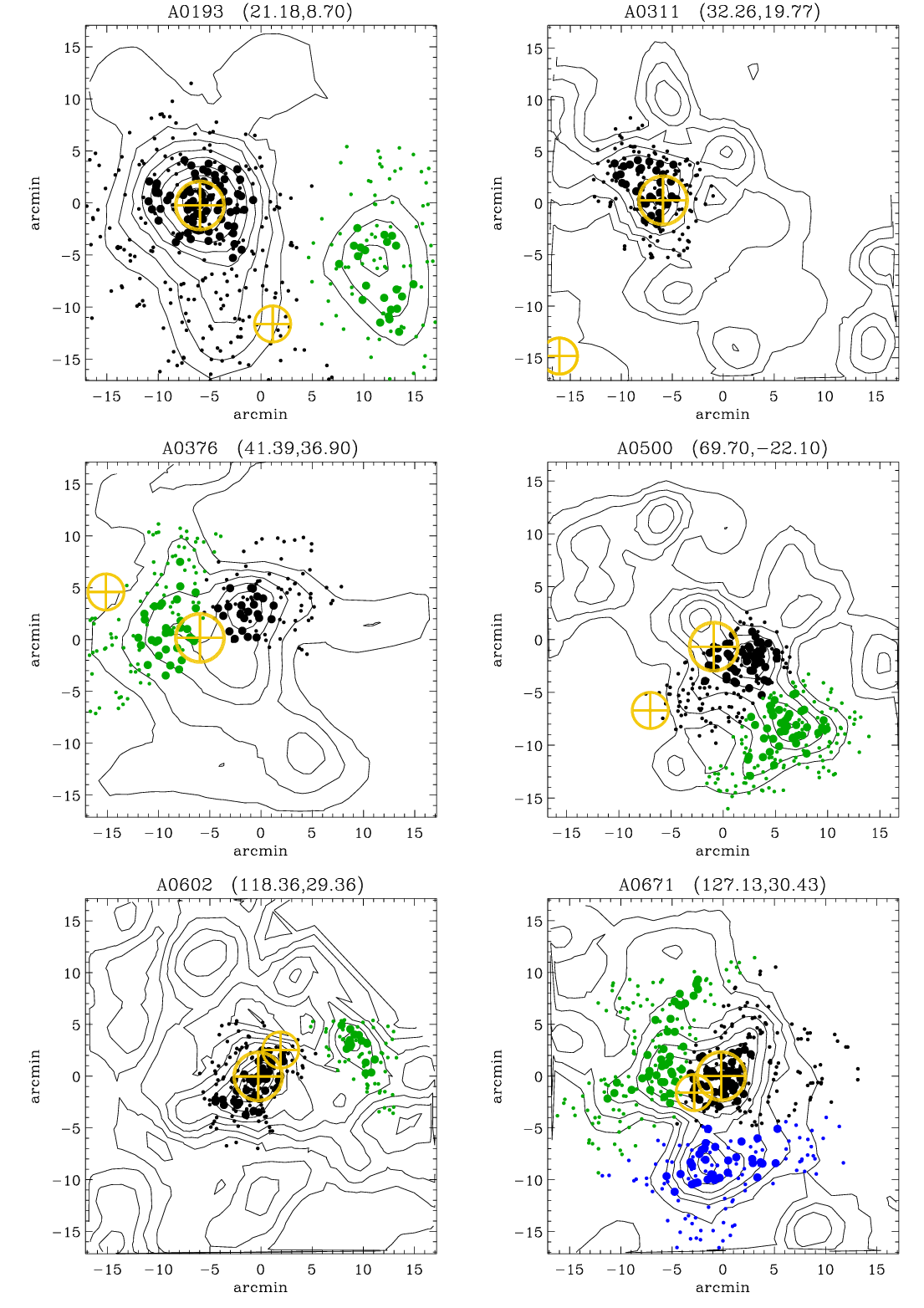}} 
\end{minipage}
\caption{(continued)}
\end{figure*} 

\addtocounter{figure}{-1} 
\begin{figure*}  \centering
\begin{minipage}{0.9\textwidth}
\resizebox{\hsize}{!}{\includegraphics{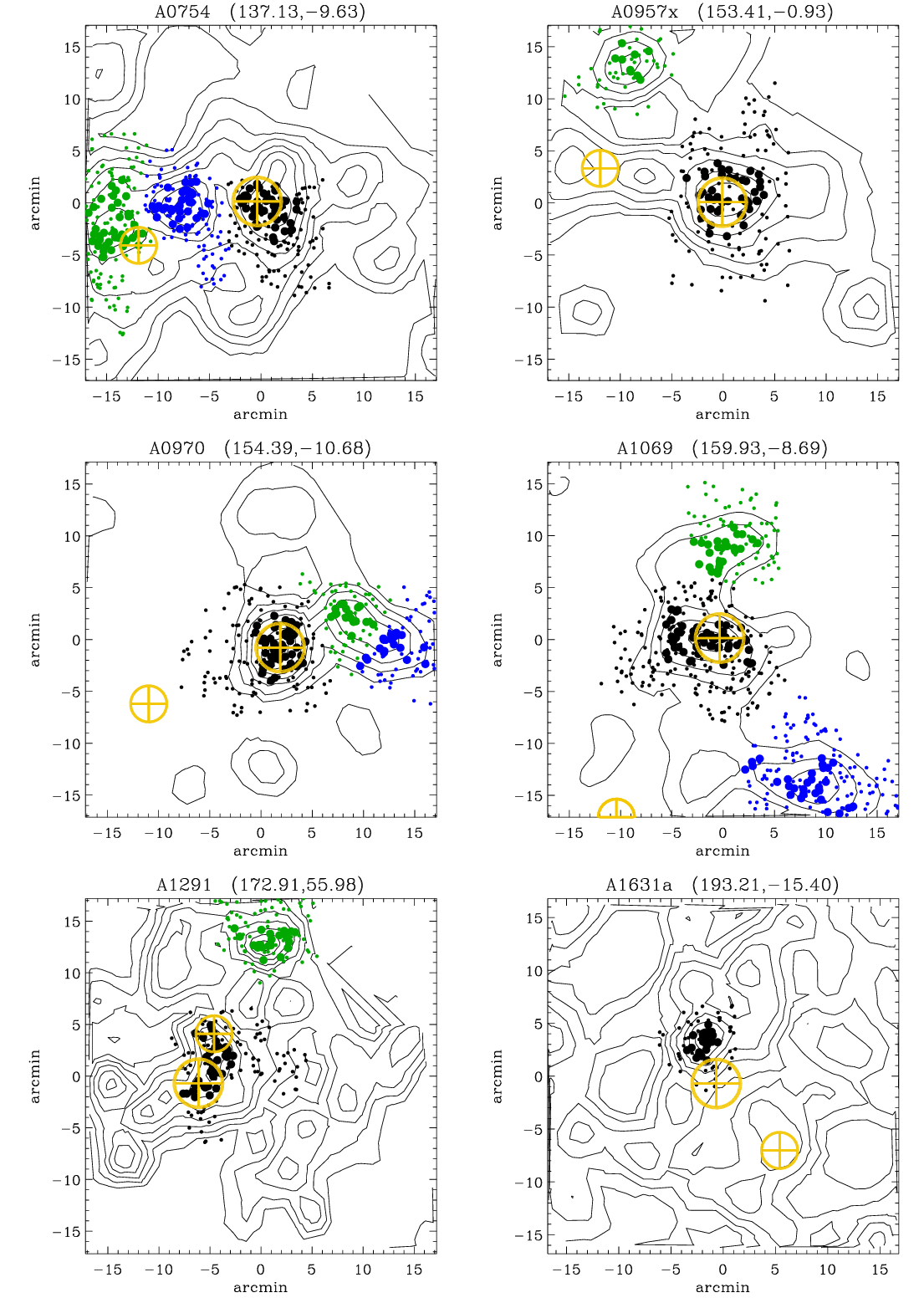}} 
\end{minipage}
\caption{(continued)}
\end{figure*} 

\addtocounter{figure}{-1} 
\begin{figure*}  \centering
\begin{minipage}{0.9\textwidth}
\resizebox{\hsize}{!}{\includegraphics{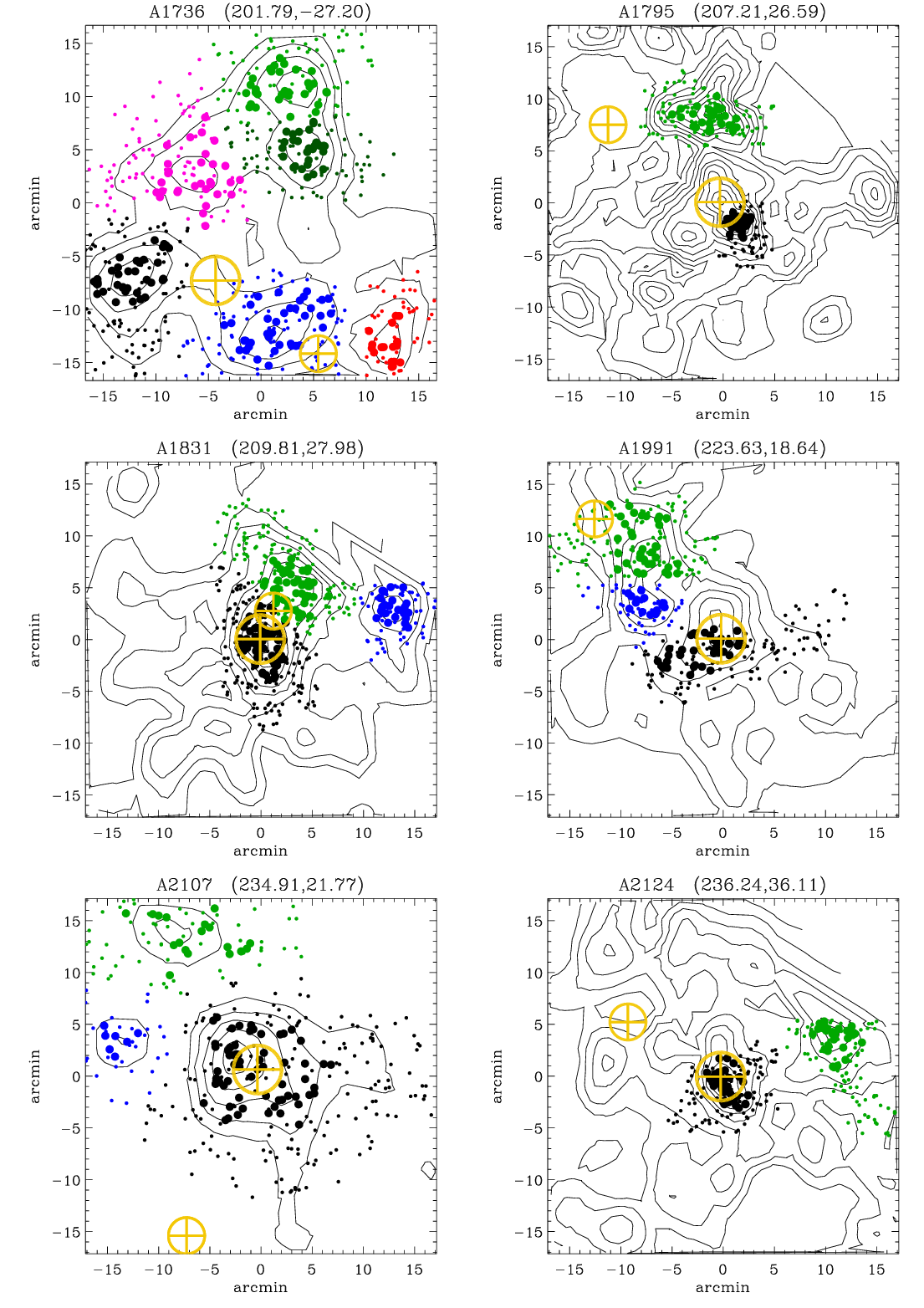}} 
\end{minipage}
\caption{(continued)}
\end{figure*} 

\addtocounter{figure}{-1} 
\begin{figure*}  \centering
\begin{minipage}{0.9\textwidth}
\resizebox{\hsize}{!}{\includegraphics{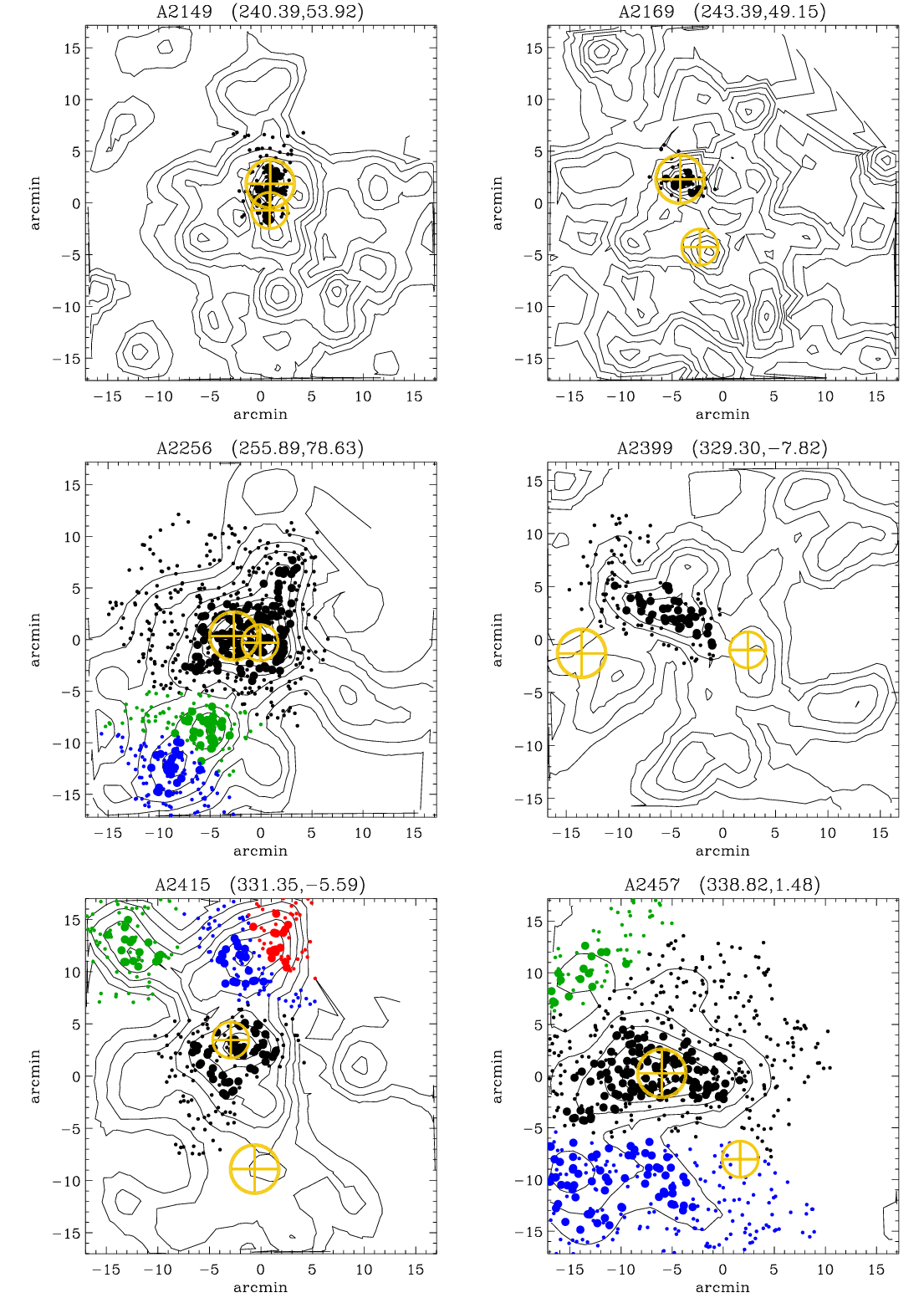}} 
\end{minipage}
\caption{(continued)}
\end{figure*} 

\addtocounter{figure}{-1} 
\begin{figure*}  \centering
\begin{minipage}{0.9\textwidth}
\resizebox{\hsize}{!}{\includegraphics{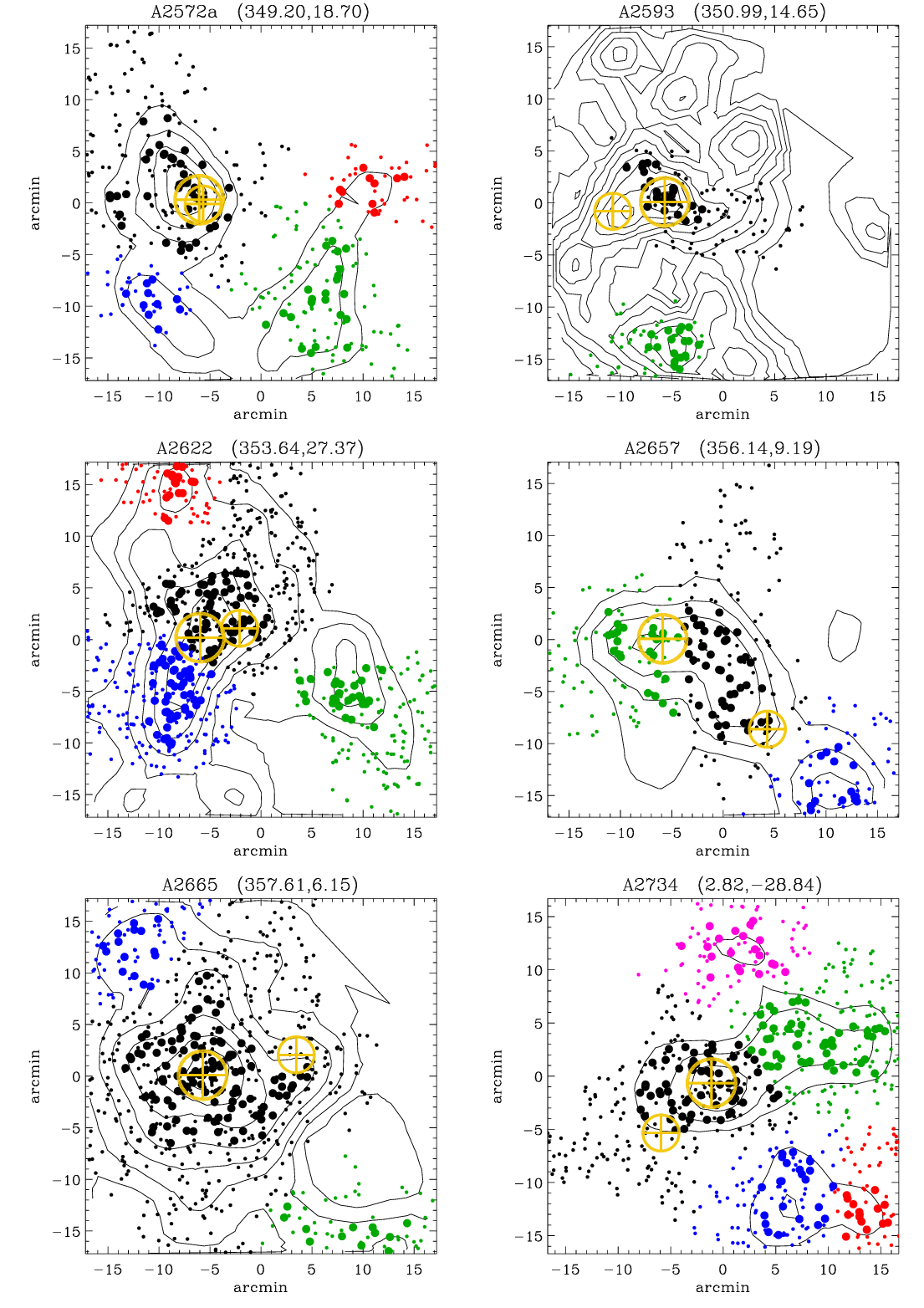}} 
\end{minipage}
\caption{(continued)}
\end{figure*} 

\addtocounter{figure}{-1} 
\begin{figure*}  \centering
\begin{minipage}{0.9\textwidth}
\resizebox{\hsize}{!}{\includegraphics{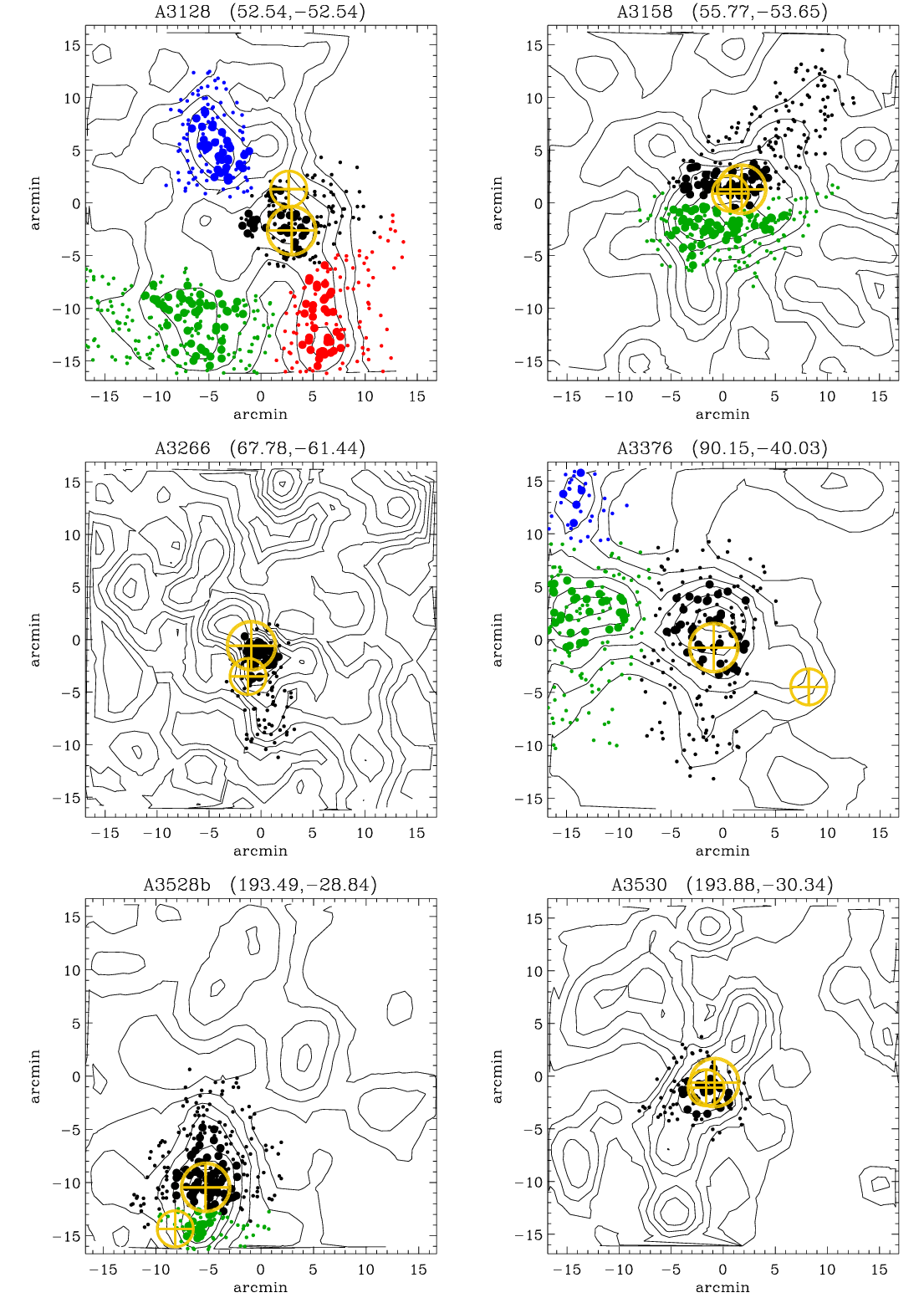}} 
\end{minipage}
\caption{(continued)}
\end{figure*} 

\addtocounter{figure}{-1} 
\begin{figure*}  \centering
\begin{minipage}{0.9\textwidth}
\resizebox{\hsize}{!}{\includegraphics{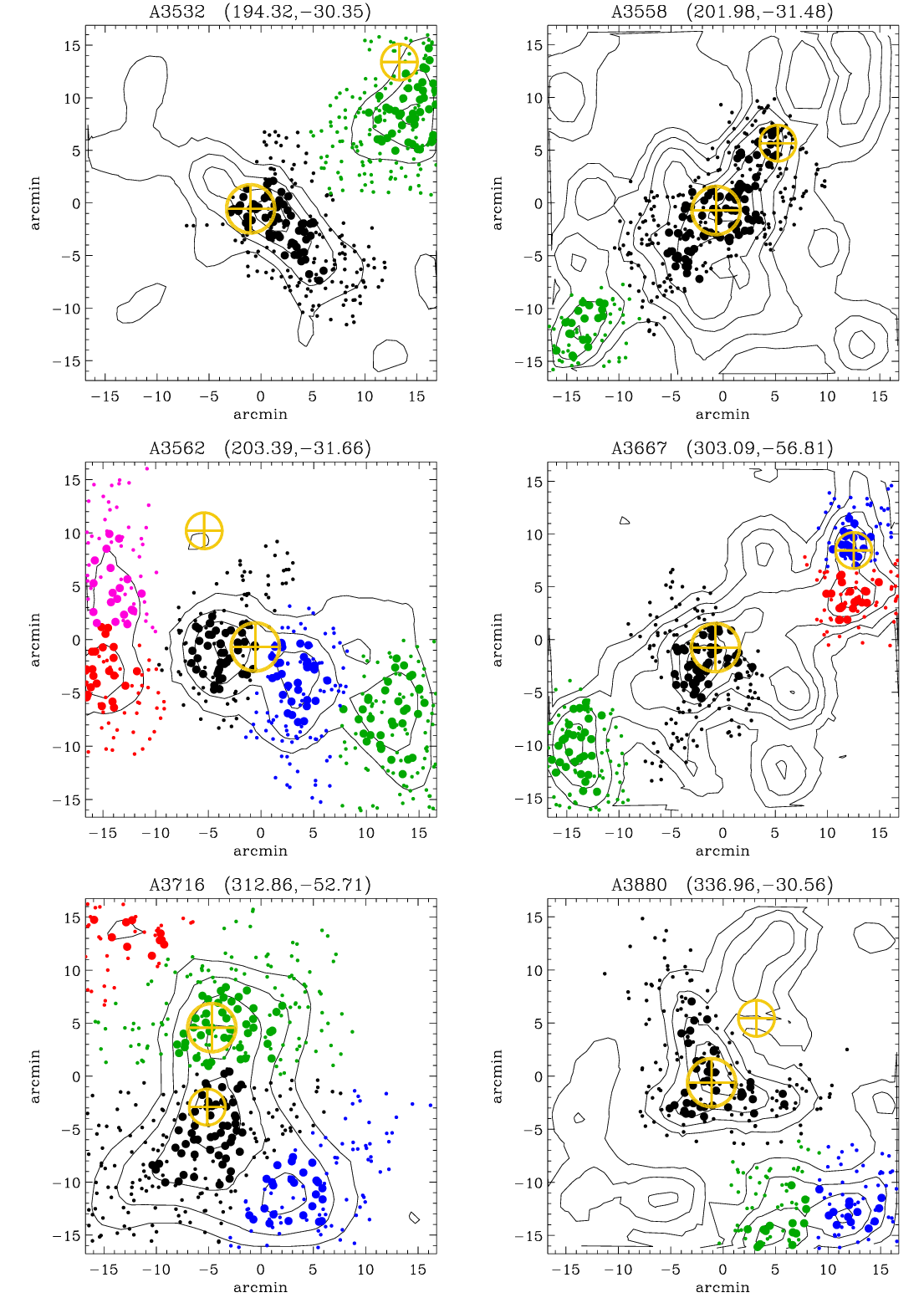}} 
\end{minipage}
\caption{(continued)}
\end{figure*} 

\addtocounter{figure}{-1} 
\begin{figure*}  \centering
\begin{minipage}{0.95\textwidth}
\resizebox{\hsize}{!}{\includegraphics{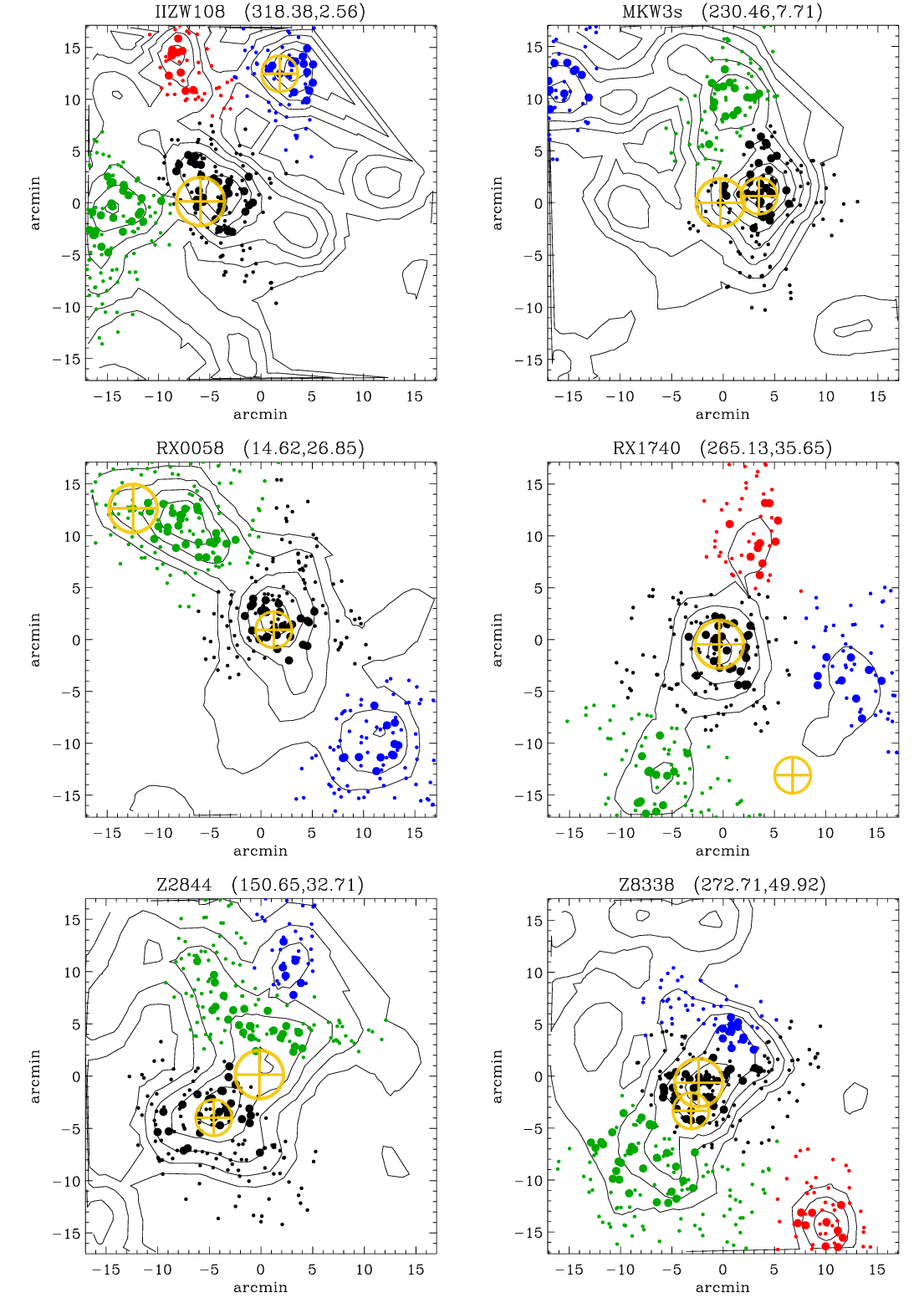}} 
\end{minipage}
\caption{(continued)}
\end{figure*} 

\addtocounter{figure}{-1} 
\begin{figure*}  \centering
\begin{minipage}{0.95\textwidth}
\resizebox{\hsize}{!}{\includegraphics{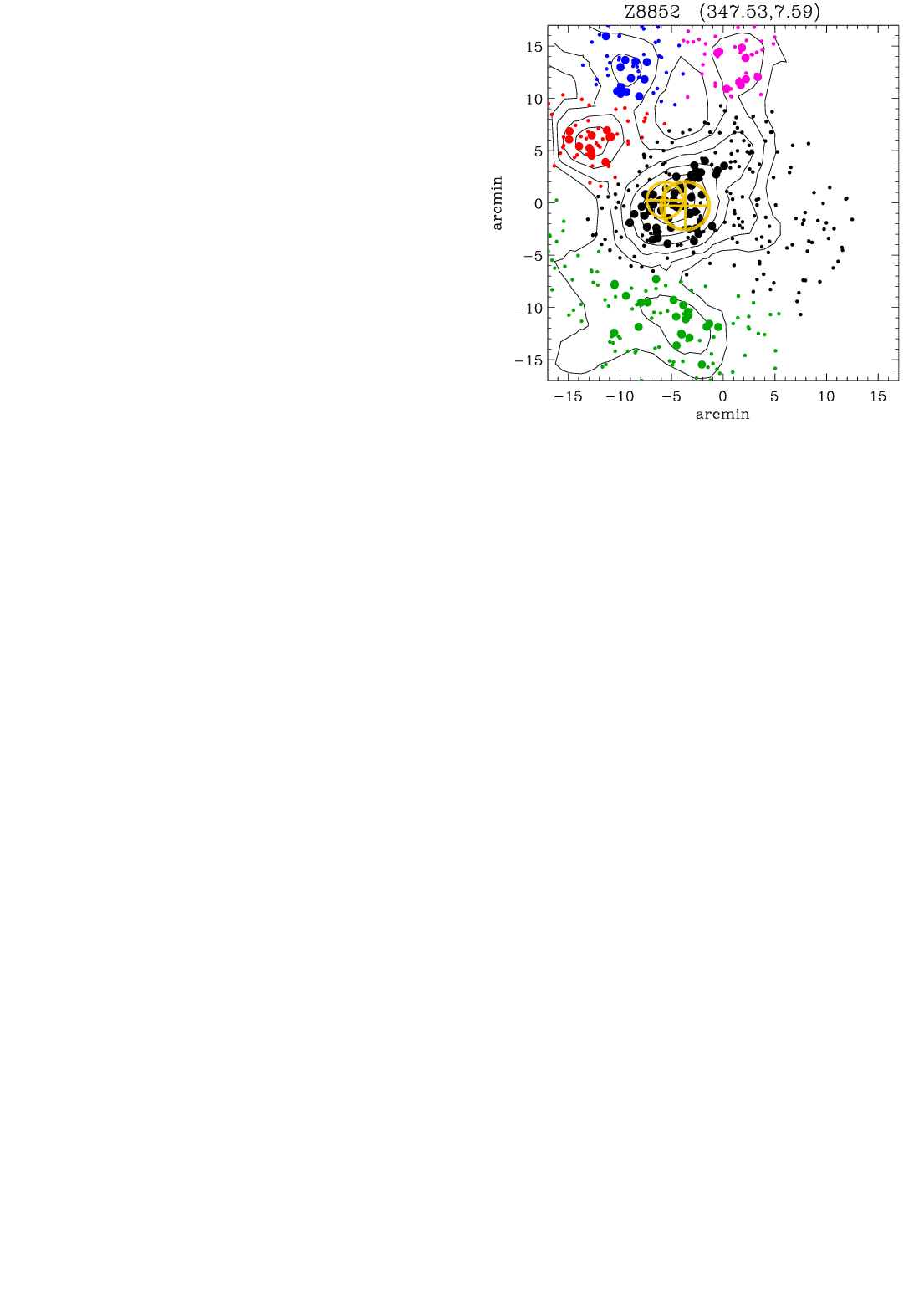}} 
\end{minipage}
\caption{(continued)}
\end{figure*} 

}

We make available contour plots of the number density fields of all
clusters in Fig.~\ref{fig:online} of the electronic version of this
Journal. In Fig.~\ref{fig:a0085} we show an example of these
plots. Isodensity contours are drawn at ten logarithmic
intervals. Galaxies belonging to the systems detected by DEDICA are
shown as dots of different colors.  We use large  symbols for brighter
galaxies ($M_V \leq -17.0$) that lie where local densities are higher than
the median local density of the structure the galaxy belongs to.
We also mark with open symbols the positions
of the first- and second-ranked cluster galaxies, BCG1 and BCG2
respectively. Color coding is black, light green, blue, red, magenta,
dark green for the main system and the subsequent substructures 
ordered as in Table \ref{tab:dedica}.

We describe and analyze in detail our catalog in the next section.

\section{Properties of substructures}
\label{s:properties}

The first problem we face in order to study the statistical and physical
properties of substructures is to determine their association with the main
structure. In fact, the main structure itself has to be identified among the
structures detected by DEDICA in each frame.

In most cases it is easy to identify the main structure of a cluster
since it is located at the center of the frame and it has a high
\parchi. In two cases (A0168 and A1736) the choice of the main
structure is complicated because there are several similar structures
near the center of the frame. In these cases we select the main
structure for its highest \parchi.

At this point we limit our analysis to members of the structure that
a) have an absolute magnitude $M_V \leq -17$ (corrected for Galactic
absorption) and that b) are in the upper half of the distribution of
DEDICA-defined local galaxy densities of the system they belong to.
The galaxy density threshold we apply allows us to separate adjacent
structures whose definition becomes more uncertain at lower galaxy
density levels. The magnitude cut increases the relative weight of the
galaxies we use to evaluate the nature of structures in the CMD.

After having identified the main structure, we need to determine which
structures in the field of view of a given cluster have to be considered
background structures.  We consider a structure a physical
substructure (or subcluster) if its color-magnitude relation (CMR
hereafter) is identical, within the errors, to the CMR of the main
structure.

As a first step we define the color-magnitude relation (CMR) of the ``whole
cluster'', i.e. of galaxies in the main structure together with all other
galaxies not assigned to any structure by DEDICA. We compute the $(B -
V)$ CMR of the Coma cluster from published data \citep{Adami+06}.  Then we
keep fixed the slope of the linear CMR of Coma and shift it to the mean
redshift of the cluster. 

In order to determine that the main structure and a substructure are at
the same redshift, we evaluate the fraction of background (red)
galaxies, $f_{bg}$, that each structure has in the CMD.  If these
fractions are identical within the errors \citep{Gehrels86}, we
consider the two structures to be at the same redshift.

In practice we determine $f_{bg}$ by assigning to the background those
galaxies of a structure that are redder than a line parallel to the
CMR and vertically shifted (i.e. redwards) by 2.33 times the root-mean
square of the colors of galaxies in the CMR. We note that the
probability that a random variable is greater than 2.33 in a Gaussian
distribution is only 1\%.

The result of the selection of main structures and substructures is
the following: 40 clusters have a total of 69 substructures at the
same redshift as the main structure, only 15 clusters are left without
substructures. A total of 35 systems are found in the background.
Considering a) the number density of poor-to-rich clusters
\citep{Mazure+96,ZGHR93}, b) the average luminosity function of
clusters \citep{Yagi+02,DePropris+03}, c) the total area covered by
the 55 cluster fields, and d) the limiting apparent magnitude
corresponding to our absolute magnitude threshold $M_V=-16.0$, we
expect to find $\sim 0.5 \pm 0.2$ background systems per cluster
field, $28 \pm 11$ in total.  This estimate is consistent with the 35
background systems we find.

The fraction of clusters with subclusters (73\%) is higher than
generally found in previous investigations \citep[typically $\sim
30$\%, see, e.g.,][and references therein]{GB02,FK06,Lopes+06}. Even if
we count all undetected clusters as clusters without substructures,
this fraction only decreases to 52\% (40/77). It is however
acknowledged that the fraction of substructured clusters depends,
among other factors, on the algorithm used to detect substructures, on
the quality and depth of the galaxy catalog. For example
\citet{KBPG01} using optical and X-ray data find that the fraction of
clusters with substructures is $\geq 45$\%, \citet{Burgett+04} using a
battery of tests detect substructures in 84\% of the 25 clusters of
their sample.

Having established the ``global'' fraction of substructured clusters,
we now investigate the degree of subclustering of individual clusters,
i.e.  the distribution of the number of substructures
$N_{sub}$ we find in our sample.

\begin{figure}
\centering \resizebox{\hsize}{!}{\includegraphics{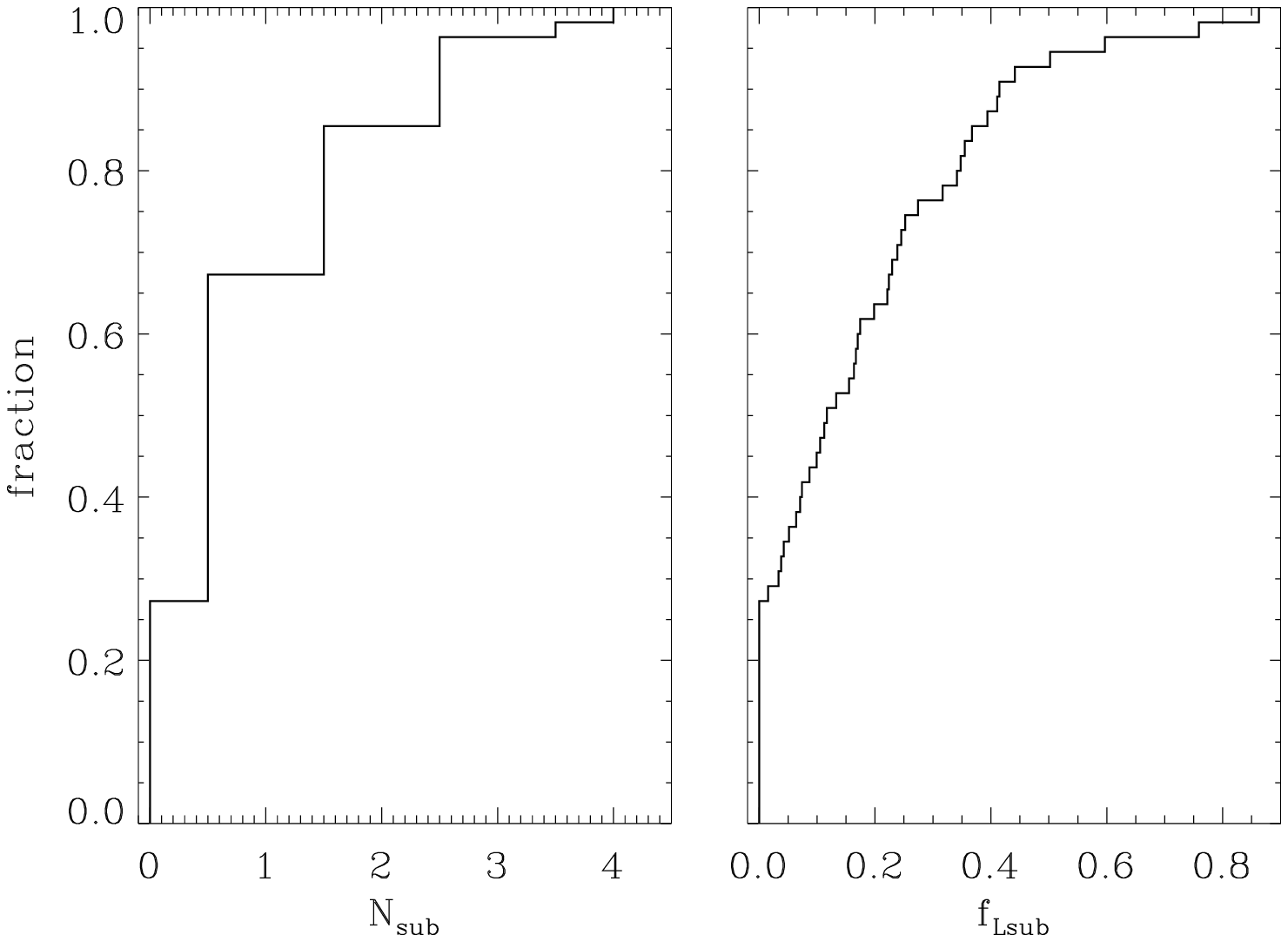}}
\caption{Cumulative distributions of the two different indicators
of subclustering: left panel  $N_{sub}$, right panel $f_{Lsub}$.}
\label{f-subdistr}
\end{figure}

We find 15 (27\%) clusters without substructures; 22 (40\%) clusters
with $N_{sub}$ = 1; 10 (18\%) clusters with $N_{sub}$ = 2; 6 (11\%)
clusters with $N_{sub}$ = 3; and 2 (3\%) clusters with $N_{sub}$ = 4.
We plot in the left panel of Fig.~\ref{f-subdistr} the integral
distribution of $N_{sub}$.

The distribution of the level of subclustering does not change when we
measure it as the fractional luminosity of subclusters, $f_{Lsub}$,
relative to the luminosity of the whole cluster (see Fig.~\ref{f-subdistr},
right panel).  The luminosities we estimate are background corrected
using the counts of \citet{Berta+06}. We use the 
ellipses output from DEDICA (see previous section)
as a measure of the area of subclusters.

We find that  $N_{sub}$ and  $f_{Lsub}$ are clearly correlated according 
to the Spearman rank-correlation test.

\begin{figure}
\centering \resizebox{\hsize}{!}{\includegraphics{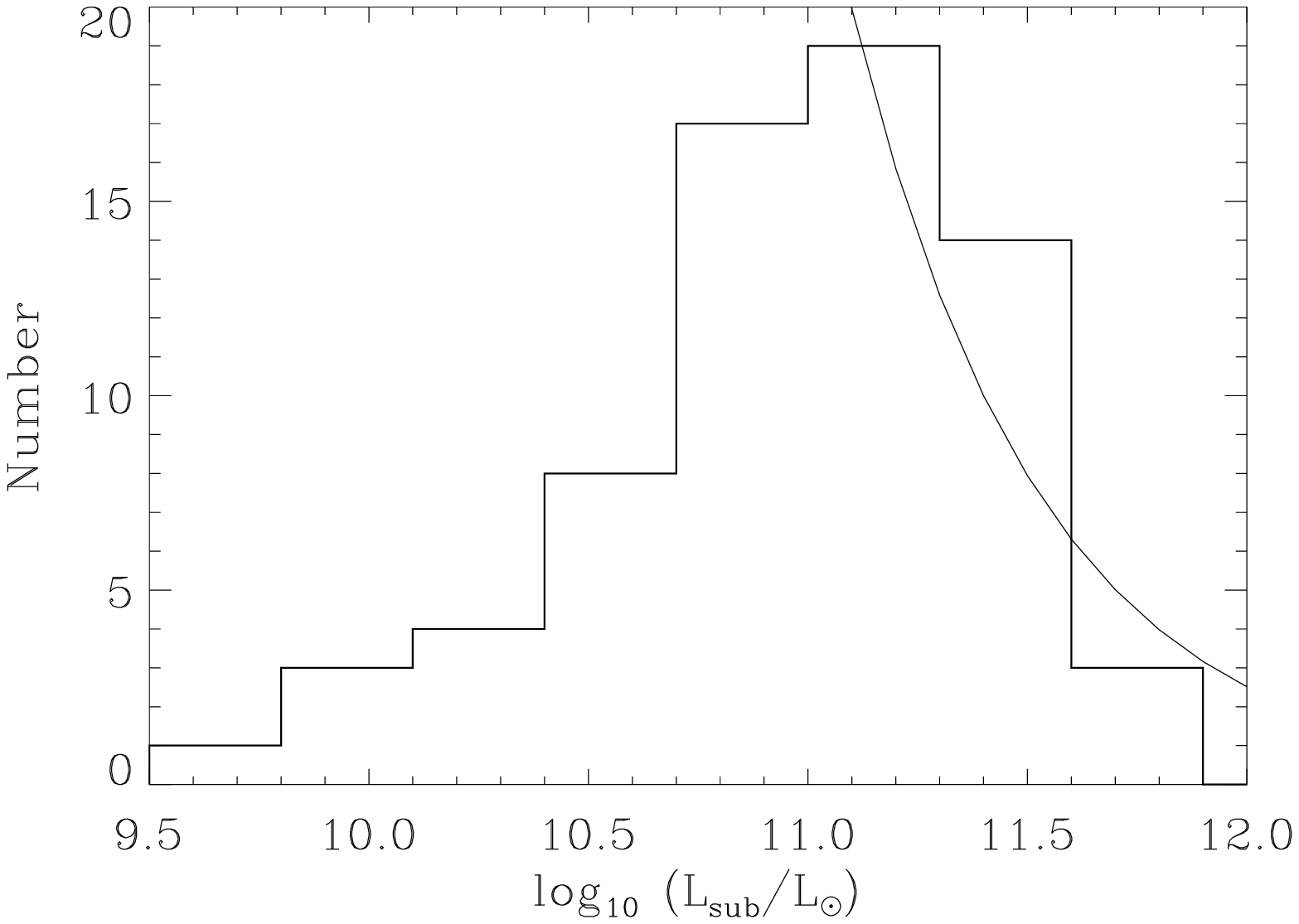}}
\caption{Observed differential distribution of subcluster luminosities (histogram) and
theoretical model \citep[arbitrary scaling; ][]{DeLucia+04}.}
\label{f-lfsubcl}
\end{figure}

We now consider the distribution of subcluster luminosities and plot the
corresponding histogram in Fig.~\ref{f-lfsubcl}.  In the same figure we also
plot with arbitrary scaling the power-law $\propto L^{-1}$. This relation is
the prediction for the differential mass function of substructures in the
cosmological simulations of \citet{DeLucia+04}.

Our observations are consistent to within the uncertainties with the
theoretical prediction of \citet{DeLucia+04} down to $L \sim 10^{11.2} \,
L_{\odot}$. The disagreement at lower luminosity is expected since: a) below
this limit galaxy-sized halos become important among the simulated
substructures, and b) only above this limit we expect our catalog to be
complete. In fact only subclusters with luminosities brighter than $L =
10^{11.2} \, L_{\odot}$ have always richnesses that are $\geq 1/3$ of the main
structure. This richness limit approximately corresponds to the completeness
limit of DEDICA detections according to our simulations (see
Sect.~\ref{s:sims}).

\section{Brightest Cluster Galaxies}
\label{s:bcg}

Here we investigate the relation between BCGs and cluster structures.

We find that, on average, BCG1s are located close to the density peak
of the main structures. In projection on the sky, the biweight average
\citep[see][]{BFG90}
distance of BCG1s from the peak of the main system is $72 \pm 11$
kpc. If we only consider the 44 BCG1s that are on the CMR and are
assigned to main systems by DEDICA, the average distance decreases to $56
\pm 8$ kpc. The fact that BCG1s are close to the center of the system is
consistent with current theoretical view on the formation of BCGs
\citep[e.g.][]{Dubinski98,Nipoti+04}.

BCG2s are more distant than BCG1s from the peak of the main system:
the biweight average distance is $345 \pm 47$ kpc. If we only consider
the 26 BCG2s that are on the CMR and are assigned to main systems by
DEDICA, the average distance decreases to $161 \pm 34$ kpc.

\begin{figure}
\centering \resizebox{\hsize}{!}{\includegraphics{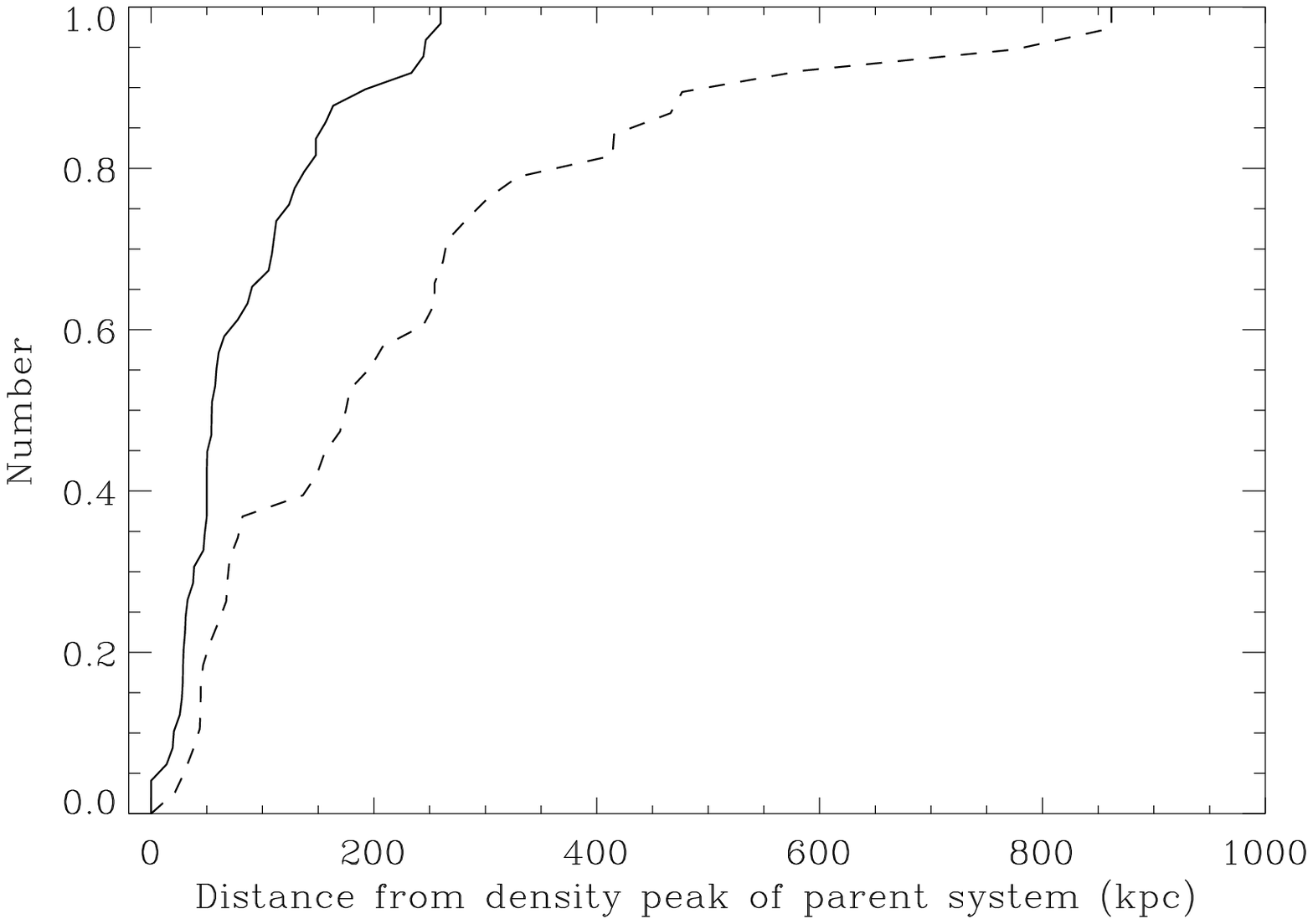}}
\caption{Cumulative distributions of distances of BCG1 (solid line)
and BCG2 (dashed line) from the density peak of their
system.}
\label{f-bcgdist}
\end{figure}

Projected distances of BCG2s from density peaks remain larger than
those of BCG1s even when we consider the density peak of the structure
or substructure they belong to.  In Fig.~\ref{f-bcgdist} we plot the
cumulative distributions of the distances of BCG1s (solid line) and
BCG2s (dashed line) from the density peak of their systems. The
distributions are different at the $>99.99$\% level according to a
Kolmogorov-Smirnov test (KS-test).

\begin{figure}
\centering \resizebox{\hsize}{!}{\includegraphics{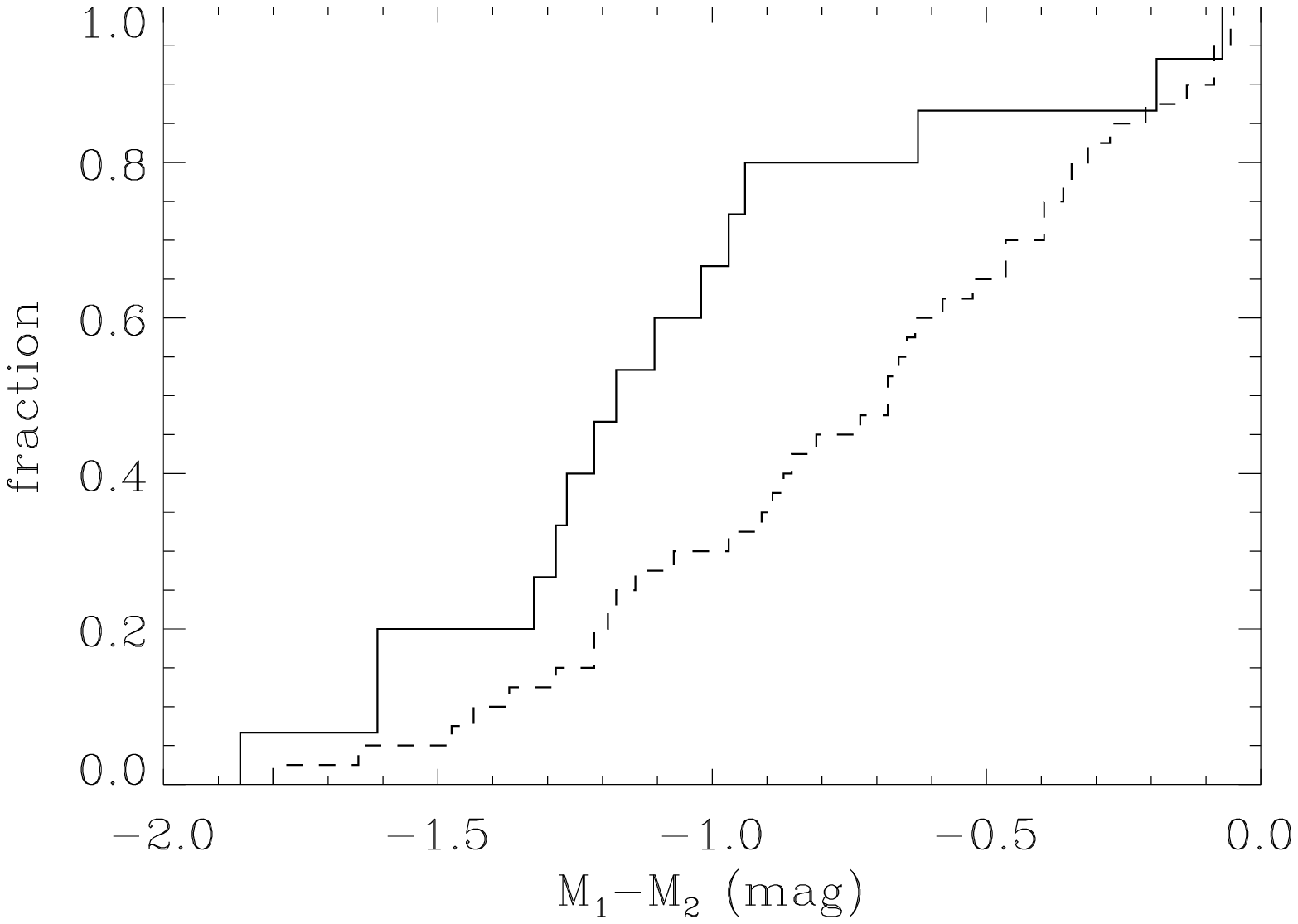}}
\caption{Cumulative distributions of the magnitude difference
between BCG1 and BCG2 in clusters with (dashed line) and
without subclusters (solid line).}
\label{f-dmag12}
\end{figure}

Now we turn to luminosities and find that the magnitude difference between
BCG1s and BCG2s, $\Delta M_{12}$, is larger in clusters without
substructures than in clusters with substructures.  In
Fig.~\ref{f-dmag12} we plot the cumulative distributions of $\Delta
M_{12}$ for clusters with (dashed line) and without (solid line)
subclusters.  The two distributions are different according to a
KS-test at the 99.1\% confidence level.  We note that \citet{LM04}
find that $\Delta M_{12}$ is independent of cluster properties. These
authors however do not consider subclustering.

In order to determine whether the higher values of $\Delta M_{12}$ in
clusters without subclusters are due to an increased luminosity of the
BCG1 ($L_1$) or to a decreased luminosity of the BCG2 ($L_2$), we
consider the luminosity of the $10^{th}$ brightest galaxy ($L_{10}$)
as a reference. The biweight 
average luminosity ratios are $<L_1/L_{10}>=8.6 \pm 1.0$ and
$<L_2/L_{10}>=3.3 \pm 0.3$ in clusters without substructures, and
$<L_1/L_{10}>=7.1 \pm 0.4$ and $<L_2/L_{10}>=3.4 \pm 0.2$ in clusters
with substructures. We then conclude that the $\Delta M_{12}$-effect
is caused by a brightening of the BCG1 relative to the BCG2 in
clusters without substructures.

The fact that $\Delta M_{12}$ is higher in clusters without substructures can
be interpreted, at least qualitatively, in the framework of the hierarchical
scenario of structure evolution. Clusters without substructures are likely to
be evolved after several merger phases. Their BCG1s have already had time to
accrete many galaxies, in particular the more massive ones, which slow down and
sink to the cluster center as the result of dynamical friction. Some of these
galaxies may even have been BCGs of the merging structures. The simulations by
\citet{DLB06} show that the BCG1s continue to increase their mass
via cannibalism even at recent times, and that there is a large variance in the
mass accretion history of BCG1s from cluster to cluster. The result of such a
cannibalism process is an increase of the BCG1 luminosity with respect to other
cluster galaxies, and in extreme cases may lead to the formation of fossil
groups \citep{KPJ06}.

However, according to these simulations, only 15\% of all BCG1s have accreted
$>$ 30\% of their mass over the last 2 Gyr, while another 15\% have accreted $<
$3\% of their mass over the same period. Our results indicate that about 60\%
of the BCG1s are more than 1 magnitude brighter than the corresponding BCG2s.
Given the size and generality of the luminosity differences it would seem that
cannibalism alone, even if present along the merging history of a given cluster,
cannot account for it. Most of the BCG1s should have then been assembled in
early times, as pointed out in the downsizing scenario for galaxy formation
\citep{Cowie+96} and entered that merging history already
with luminosity not far form the present one. 

\section{Summary}
\label{s:summary}
In this paper we search for and characterize cluster substructures, or
subclusters, in the sample of 77 nearby clusters of the WINGS
\citep{Fasano+06}.  This sample is an almost complete sample in X-ray
flux in the redshift range $0.04<z<0.07$.

We detect substructures in the spatial projected distribution of
galaxies in the cluster fields using DEDICA \citep{Pisani93,Pisani96}
an adaptive-kernel technique. DEDICA has the following advantages for
our study of WINGS clusters:

a) DEDICA gives a total description of the clustering pattern, in
particular membership probability and significance of structures besides
geometrical properties.

b) DEDICA is scale invariant

c) DEDICA does not assume any property of the clusters, i.e. it is completely
non-parametric. In particular it does not require particularly rich samples to
run effectively.

In order to test DEDICA and to set guidelines for the interpretation 
of the results of the application of DEDICA to our observations 
we run DEDICA on several sets of simulated fields containing a cluster 
with and without subclusters. 

We find that: a) DEDICA always identifies both cluster and 
subcluster even when the substructure richness ratio cluster-to-subcluster
is \rsc = 8, b) DEDICA  recovers a large
fraction  of members, almost irrespective of the richness of the original
structure ($\ga 70\%$ in most cases), c) structures with richness
ratios \rsc  $\la 3$ are always distinguishable from noise fluctuations of
the poissonian simulated field.

These simulations also allow us to define a threshold that we use 
to identify significant structures in the observed fields. 

We apply our clustering procedure to the 77 clusters of the WINGS
sample. We cut galaxy catalogs to the absolute magnitude threshold
$M_V = -16.0$ in order to maximize the signal-to-noise ratio of the
detected subclusters.

We detect at least one significant structure in 55 (71\%) cluster
fields. We find that 12 clusters (16\%) have no structure above the
threshold (undetected).  In the remaining 10 (13\%) clusters we find
significant structures only at the border of the field of view. In
absence of a detection in the center of the frame, we consider these
border structures unrelated to the target cluster.  We also verify
that in the CMD these border structures are redder
than expected given the redshift of the target cluster.  We consider
also these clusters undetected.

We provide the coordinates of all substructures in the 55 clusters
together with their main properties.

Using the CMR of the early-type cluster galaxies
we separate "true" subclusters from unrelated background structures. 
We find that 40 clusters out of 55 (73\%) have a total of 69 substructures 
with 15 clusters left without substructures.

The fraction of clusters with subclusters (73\%) we identify is higher
than most previously published values \citep[typically $\sim 30$\%, see,
e.g.,][and references therein]{GB02}. It is however acknowledged that
the fraction of substructured clusters depends, among other factors,
on the algorithm used to detect substructures, on the quality and
depth of the galaxy catalog \citep{KBPG01,Burgett+04}.

Another important result of our analysis is the distribution of
subcluster luminosities. In the luminosity range where our
substructure detection is complete ($L \geq 10^{11.2} \, L_{\odot}$),
we find that the distribution of subcluster luminosities is in
agreement with the power-law $\propto L^{-1}$ predicted for the
differential mass function of substructures in the cosmological
simulations of \citet{DeLucia+04}.

Finally, we investigate the relation between BCGs and cluster structures.

We find that, on average, BCG1s are located close to the density peak
of the main structures. In projection on the sky, the biweight average
distance of BCG1s from the peak of the main system is $72 \pm 11$
kpc. BCG2s are significantly more distant than BCG1s from the peak of the
main system ($345 \pm 47$ kpc). 

The fact that BCG1s are close to the center of the system is consistent
with current theoretical view on the formation of BCGs \citep{Dubinski98}.

A more surprising result is that the magnitude difference between
BCG1s and BCG2s, $\Delta M_{12}$, is significantly larger in clusters without
substructures than in clusters with substructures. This fact 
may be interpreted in the framework of the hierarchical
scenario of structure evolution \citep[e.g.][]{DLB06}. 

\bibliography{babbage}
 
\clearpage
\onecolumn

\begin{appendix}
\section{The catalog of substructures}
\label{app:table}
We provide here the catalog of substructures.  In
Table~\ref{tab:dedica} we give, for each substructure: (1) the name of
the parent cluster; (2) the classification of the structure as main
(M), subcluster (S), or background (B) together with their order
number; (3) right ascension (J2000), and (4) declination (J2000) in
decimal degrees of the DEDICA peak; the parameters of the ellipse we
obtain from the variance matrix of the coordinates of galaxies in the
substructure, i.e.  (5) major axis in arcminutes, (6) ellipticity, and
(7) position angle in degrees; (8) luminosity; (9) \parchi. 

\begin{longtable}{lcrrccrcc}\hline\hline
\hline
~~~ID & class & $\alpha_{J2000}$~~ & $\delta_{J2000}$~~ & a  & e & PA~ & $ L $ & \parchi \\
  & & (deg)~~~ & (deg)~~~ & (arcmin) & & (deg) & $(10^{12} \, L_{\odot})$ & \\
\hline
\endhead
A0085 & M & 10.4752 & -9.3025 & 2.0 & 0.23 & -17. & 0.41536 & 48.4 \\
A0085 & S1 & 10.4410 & -9.4430 & 1.8 & 0.35 & -39. & 0.17649 & 42.9 \\
A0085 & S2 & 10.3947 & -9.3501 & 2.3 & 0.40 & -72. & 0.12337 & 32.8 \\
A0119 & M & 14.0625 & -1.2630 & 4.1 & 0.44 & -65. & 0.83955 & 63.4 \\
A0119 & S1 & 14.1183 & -1.2106 & 4.6 & 0.60 & -23. & 0.26847 & 50.8 \\
A0119 & S2 & 14.0267 & -1.0441 & 3.4 & 0.34 & 80. & 0.03592 & 32.0 \\
A0119 & B1 & 13.9402 & -1.4979 & 3.1 & 0.39 & 46. & -- & 23.5 \\
A0147 & M & 17.0648 & 2.2033 & 3.9 & 0.45 & 79. & 0.31392 & 45.2 \\
A0147 & S1 & 16.8673 & 2.1393 & 4.1 & 0.25 & -50. & 0.05638 & 24.8 \\
A0147 & S2 & 17.1925 & 1.9284 & 4.4 & 0.38 & 55. & 0.05052 & 21.4 \\
A0147 & B1 & 17.0753 & 2.3174 & 4.4 & 0.37 & 75. & -- & 58.0 \\
A0151 & M & 17.2186 & -15.4219 & 1.7 & 0.26 & -16. & 0.47344 & 39.9 \\
A0151 & S1 & 17.3516 & -15.3652 & 2.1 & 0.37 & -58. & 0.13761 & 42.9 \\
A0151 & S2 & 17.2632 & -15.5564 & 1.6 & 0.26 & -53. & 0.19762 & 40.9 \\
A0151 & B1 & 17.1375 & -15.6116 & 1.5 & 0.08 & -4. & -- & 59.0 \\
A0160 & M & 18.2344 & 15.5126 & 3.6 & 0.37 & 82. & 0.55525 & 66.7 \\
A0160 & S1 & 18.2483 & 15.3138 & 5.0 & 0.41 & 85. & 0.03120 & 38.1 \\
A0160 & S2 & 18.1141 & 15.7501 & 3.0 & 0.16 & 86. & 0.15196 & 28.3 \\
A0160 & S3 & 17.9981 & 15.4150 & 3.9 & 0.41 & 0. & 0.06315 & 27.8 \\
A0168 & M & 18.7755 & 0.3999 & 3.1 & 0.32 & -11. & 0.24492 & 30.5 \\
A0168 & S1 & 18.8799 & 0.2993 & 2.0 & 0.33 & 4. & 0.06871 & 28.6 \\
A0193 & M & 21.2894 & 8.6994 & 2.1 & 0.08 & 36. & 0.61982 & 105.7 \\
A0193 & B1 & 20.9945 & 8.6119 & 4.9 & 0.45 & -1. & -- & 39.1 \\
A0311 & M & 32.3793 & 19.7722 & 2.3 & 0.19 & 43. & 0.43320 & 44.0 \\
A0376 & M & 41.4276 & 36.9517 & 1.7 & 0.07 & -67. & 0.13477 & 40.8 \\
A0376 & S1 & 41.5569 & 36.9214 & 4.4 & 0.49 & -22. & 0.24350 & 33.6 \\
A0500 & M & 69.6476 & -22.1308 & 2.0 & 0.31 & 16. & 0.41203 & 45.5 \\
A0500 & S1 & 69.5915 & -22.2377 & 2.3 & 0.19 & 36. & 0.20520 & 47.3 \\
A0602 & M & 118.3638 & 29.3528 & 1.8 & 0.55 & -46. & 0.20112 & 55.4 \\
A0602 & S1 & 118.1848 & 29.4145 & 2.3 & 0.52 & 31. & 0.08470 & 34.8 \\
A0671 & M & 127.1237 & 30.4269 & 1.6 & 0.24 & -51. & 0.68582 & 69.8 \\
A0671 & S1 & 127.2241 & 30.4342 & 2.0 & 0.40 & -5. & 0.19736 & 44.3 \\
A0671 & S2 & 127.1617 & 30.2967 & 1.9 & 0.24 & -90. & 0.13778 & 43.0 \\
A0754 & M & 137.1073 & -9.6370 & 2.0 & 0.25 & 53. & 0.56063 & 46.8 \\
A0754 & S1 & 137.3707 & -9.6760 & 3.2 & 0.53 & -8. & 0.30590 & 54.9 \\
A0754 & S2 & 137.2619 & -9.6367 & 1.7 & 0.14 & 76. & 0.23734 & 51.2 \\
A0957x & M & 153.4095 & -0.9259 & 2.0 & 0.09 & -83. & 0.42106 & 38.6 \\
A0957x & B1 & 153.5517 & -0.7023 & 2.2 & 0.44 & -63. & -- & 37.9 \\
A0970 & M & 154.3595 & -10.6921 & 1.5 & 0.27 & -30. & 0.46130 & 62.5 \\
A0970 & S1 & 154.2369 & -10.6422 & 1.7 & 0.15 & 15. & 0.13660 & 42.3 \\
A0970 & B1 & 154.1833 & -10.6771 & 1.8 & 0.23 & -76. & -- & 32.2 \\
A1069 & M & 159.9418 & -8.6883 & 2.8 & 0.31 & 52. & 0.37270 & 50.2 \\
A1069 & S1 & 159.9286 & -8.5506 & 2.4 & 0.23 & 88. & 0.18532 & 32.7 \\
A1069 & B1 & 159.7678 & -8.9262 & 3.5 & 0.55 & 77. & -- & 54.7 \\
A1291 & M & 173.0467 & 56.0255 & 2.5 & 0.51 & -11. & 0.25272 & 32.1 \\
A1291 & S1 & 172.9090 & 56.1872 & 1.4 & 0.48 & -82. & 0.03530 & 37.6 \\
A1631a & M & 193.2410 & -15.3413 & 1.4 & 0.35 & 40. & 0.20077 & 33.9 \\
A1736 & M & 202.0097 & -27.3131 & 3.1 & 0.35 & 58. & 0.41824 & 52.1 \\
A1736 & S1 & 201.7305 & -27.0170 & 2.8 & 0.32 & 9. & 0.24023 & 42.6 \\
A1736 & S2 & 201.7662 & -27.4067 & 3.4 & 0.28 & 7. & 0.14528 & 42.3 \\
A1736 & S3 & 201.5672 & -27.4291 & 2.7 & 0.44 & 73. & 0.16926 & 40.4 \\
A1736 & S4 & 201.9057 & -27.1600 & 3.5 & 0.21 & -1. & 0.24192 & 32.9 \\
A1736 & S5 & 201.7036 & -27.1236 & 3.0 & 0.39 & -12. & 0.40395 & 31.7 \\
A1795 & M & 207.1911 & 26.5586 & 0.6 & 0.17 & 55. & 0.12341 & 52.4 \\
A1795 & S1 & 207.2329 & 26.7362 & 1.3 & 0.38 & 82. & 0.05123 & 46.7 \\
A1831 & M & 209.8120 & 27.9714 & 1.9 & 0.43 & 9. & 1.08418 & 56.0 \\
A1831 & S1 & 209.7356 & 28.0636 & 2.1 & 0.34 & 41. & 0.36295 & 59.7 \\
A1831 & B1 & 209.5725 & 28.0206 & 1.7 & 0.25 & -10. & -- & 47.7 \\
A1991 & M & 223.6405 & 18.6390 & 2.3 & 0.54 & -78. & 0.28195 & 40.3 \\
A1991 & S1 & 223.7575 & 18.7812 & 2.5 & 0.32 & 41. & 0.11412 & 49.9 \\
A1991 & B1 & 223.7683 & 18.7022 & 1.7 & 0.31 & 71. & -- & 36.6 \\
A2107 & M & 234.9497 & 21.8075 & 2.7 & 0.19 & 48. & 0.50994 & 61.1 \\
A2107 & B1 & 235.0699 & 22.0127 & 4.3 & 0.48 & 83. & -- & 32.9 \\
A2107 & B2 & 235.1409 & 21.8276 & 2.4 & 0.10 & 55. & -- & 20.4 \\
A2124 & M & 236.2400 & 36.0990 & 1.3 & 0.24 & 32. & 0.41727 & 43.3 \\
A2124 & B1 & 236.0207 & 36.1779 & 1.6 & 0.22 & 34. & -- & 59.7 \\
A2149 & M & 240.3723 & 53.9406 & 1.5 & 0.46 & -10. & 0.37347 & 48.7 \\
A2169 & M & 243.4867 & 49.1875 & 0.6 & 0.22 & 72. & 0.15358 & 34.2 \\
A2256 & M & 255.9260 & 78.6412 & 1.9 & 0.29 & -86. & 1.46563 & 95.6 \\
A2256 & B1 & 256.3094 & 78.4886 & 2.2 & 0.48 & 75. & -- & 48.2 \\
A2256 & B2 & 256.6024 & 78.4283 & 2.0 & 0.12 & -88. & -- & 46.8 \\
A2399 & M & 329.3693 & -7.7772 & 3.5 & 0.64 & -26. & 0.40505 & 38.8 \\
A2415 & M & 331.3829 & -5.5444 & 2.3 & 0.23 & -60. & 0.36780 & 44.8 \\
A2415 & S1 & 331.5610 & -5.3960 & 2.3 & 0.33 & 52. & 0.05032 & 33.9 \\
A2415 & B1 & 331.3800 & -5.4017 & 1.9 & 0.34 & 32. & -- & 41.4 \\
A2415 & B2 & 331.3295 & -5.3890 & 1.6 & 0.41 & 4. & -- & 37.3 \\
A2457 & M & 338.9462 & 1.4765 & 4.3 & 0.50 & -84. & 0.88720 & 107.3 \\
A2457 & S1 & 339.0392 & 1.6459 & 4.1 & 0.65 & -50. & 0.05960 & 23.1 \\
A2457 & B1 & 339.0667 & 1.3266 & 5.6 & 0.53 & 77. & -- & 73.8 \\
A2572a & M & 349.3192 & 18.7197 & 2.9 & 0.39 & 23. & 0.44749 & 47.8 \\
A2572a & S1 & 349.1122 & 18.5320 & 4.1 & 0.25 & 8. & 0.07320 & 34.5 \\
A2572a & S2 & 349.3851 & 18.5395 & 2.6 & 0.31 & 67. & 0.05345 & 25.1 \\
A2572a & S3 & 349.0037 & 18.7220 & 3.2 & 0.34 & 86. & 0.00884 & 20.5 \\
A2593 & M & 351.0766 & 14.6539 & 1.1 & 0.25 & 58. & 0.28333 & 33.8 \\
A2593 & S1 & 351.0677 & 14.4048 & 2.2 & 0.42 & 80. & 0.09810 & 27.0 \\
A2622 & M & 353.7384 & 27.3856 & 3.1 & 0.09 & 76. & 0.48920 & 68.1 \\
A2622 & S1 & 353.4880 & 27.2877 & 4.2 & 0.35 & 46. & 0.03070 & 35.1 \\
A2622 & B1 & 353.7837 & 27.3182 & 3.0 & 0.49 & -5. & -- & 53.7 \\
A2622 & B2 & 353.8009 & 27.6217 & 2.6 & 0.38 & 68. & -- & 29.9 \\
A2657 & M & 356.1725 & 9.1818 & 4.5 & 0.47 & 22. & 0.27061 & 49.6 \\
A2657 & S1 & 356.2755 & 9.1799 & 3.2 & 0.35 & 10. & 0.20771 & 34.0 \\
A2657 & B1 & 355.9569 & 8.9422 & 2.8 & 0.06 & -10. & -- & 38.6 \\
A2665 & M & 357.7050 & 6.1582 & 3.6 & 0.26 & 71. & 0.67950 & 121.8 \\
A2665 & S1 & 357.4003 & 5.8659 & 4.9 & 0.64 & 84. & 0.01780 & 17.5 \\
A2665 & B1 & 357.8218 & 6.3522 & 3.5 & 0.44 & -50. & -- & 32.7 \\
A2734 & M & 2.8363 & -28.8652 & 3.4 & 0.33 & 3. & 0.48700 & 56.3 \\
A2734 & S1 & 2.6950 & -28.7728 & 4.0 & 0.27 & -7. & 0.18970 & 55.0 \\
A2734 & S2 & 2.6987 & -29.0394 & 3.2 & 0.24 & 55. & 0.03030 & 43.3 \\
A2734 & S3 & 2.5727 & -29.0562 & 3.4 & 0.51 & 84. & 0.03100 & 33.1 \\
A2734 & B1 & 2.7701 & -28.6488 & 3.7 & 0.39 & 14. & -- & 28.0 \\
A3128 & M & 52.4825 & -52.5764 & 2.2 & 0.17 & -36. & 0.40452 & 37.2 \\
A3128 & S1 & 52.7366 & -52.7089 & 4.1 & 0.44 & -9. & 0.25240 & 51.8 \\
A3128 & S2 & 52.6655 & -52.4413 & 2.3 & 0.32 & -65. & 0.19646 & 51.0 \\
A3128 & S3 & 52.3697 & -52.7570 & 3.2 & 0.47 & 81. & 0.17169 & 39.0 \\
A3158 & M & 55.7477 & -53.6334 & 2.6 & 0.62 & 2. & 0.70205 & 52.8 \\
A3158 & S1 & 55.8382 & -53.6780 & 3.4 & 0.58 & 10. & 0.45553 & 53.4 \\
A3266 & M & 67.7893 & -61.4637 & 1.1 & 0.27 & -72. & 0.42993 & 63.5 \\
A3376 & M & 90.1628 & -39.9950 & 2.5 & 0.14 & -43. & 0.33708 & 43.1 \\
A3376 & S1 & 90.4344 & -39.9776 & 2.7 & 0.20 & -4. & 0.21279 & 59.8 \\
A3376 & S2 & 90.4712 & -39.7946 & 2.1 & 0.39 & -88. & 0.00904 & 31.2 \\
A3528b & M & 193.5928 & -29.0136 & 1.3 & 0.04 & -24. & 0.65638 & 66.4 \\
A3528b & S1 & 193.6030 & -29.0721 & 1.3 & 0.26 & 10. & 0.16706 & 59.0 \\
A3530 & M & 193.9098 & -30.3606 & 1.9 & 0.26 & 33. & 0.53043 & 34.9 \\
A3532 & M & 194.3035 & -30.3732 & 3.6 & 0.52 & -44. & 0.76920 & 51.1 \\
A3532 & B1 & 194.0413 & -30.2130 & 3.8 & 0.38 & 66. & -- & 56.3 \\
A3558 & M & 201.9587 & -31.4892 & 4.9 & 0.54 & 49. & 1.14860 & 64.1 \\
A3558 & B1 & 202.2501 & -31.6887 & 2.6 & 0.49 & 44. & -- & 37.6 \\
A3562 & M & 203.4603 & -31.6812 & 2.5 & 0.18 & 82. & 0.39087 & 42.4 \\
A3562 & S1 & 203.1622 & -31.7742 & 4.0 & 0.40 & -86. & 0.15010 & 51.1 \\
A3562 & S2 & 203.3137 & -31.6953 & 3.6 & 0.40 & 76. & 0.11706 & 41.0 \\
A3562 & S3 & 203.6982 & -31.7171 & 2.5 & 0.21 & -50. & 0.07820 & 36.7 \\
A3562 & S4 & 203.6541 & -31.5969 & 4.0 & 0.55 & -81. & 0.06542 & 30.3 \\
A3667 & M & 303.1637 & -56.8598 & 2.1 & 0.14 & 23. & 0.56803 & 42.0 \\
A3667 & S1 & 303.5297 & -56.9660 & 3.0 & 0.39 & -86. & 0.27086 & 39.2 \\
A3667 & S2 & 302.7241 & -56.6674 & 1.5 & 0.17 & 75. & 0.28948 & 34.0 \\
A3667 & S3 & 302.7081 & -56.7557 & 2.3 & 0.39 & 12. & 0.09961 & 33.7 \\
A3716 & M & 312.9910 & -52.7677 & 4.7 & 0.38 & 36. & 0.76450 & 77.9 \\
A3716 & S1 & 312.9769 & -52.6434 & 3.5 & 0.22 & -6. & 0.49159 & 56.9 \\
A3716 & B1 & 312.7735 & -52.8976 & 4.1 & 0.40 & 28. & -- & 48.4 \\
A3716 & B2 & 313.1888 & -52.4785 & 3.2 & 0.23 & 21. & -- & 22.6 \\
A3880 & M & 336.9796 & -30.5474 & 3.8 & 0.33 & -50. & 0.25840 & 44.1 \\
A3880 & B1 & 336.8684 & -30.8171 & 2.5 & 0.30 & 64. & -- & 30.9 \\
A3880 & B2 & 336.7356 & -30.7839 & 2.7 & 0.35 & 46. & -- & 28.4 \\
IIZW108 & M & 318.4443 & 2.5706 & 2.6 & 0.38 & 42. & 0.49940 & 33.4 \\
IIZW108 & S1 & 318.6247 & 2.5533 & 3.2 & 0.36 & -14. & 0.08565 & 42.3 \\
IIZW108 & B1 & 318.3335 & 2.7751 & 1.6 & 0.24 & 43. & -- & 44.5 \\
IIZW108 & B2 & 318.5190 & 2.8039 & 2.5 & 0.44 & 22. & -- & 33.6 \\
MKW3s & M & 230.3916 & 7.7281 & 2.3 & 0.30 & -8. & 0.37614 & 48.7 \\
MKW3s & S1 & 230.4576 & 7.8769 & 3.3 & 0.46 & -22. & 0.04585 & 39.3 \\
MKW3s & B1 & 230.7349 & 7.8882 & 2.0 & 0.40 & 16. & -- & 25.5 \\
RX0058 & M & 14.5875 & 26.8816 & 2.4 & 0.22 & -31. & 0.31967 & 44.2 \\
RX0058 & S1 & 14.7652 & 27.0424 & 3.8 & 0.60 & 64. & 0.31661 & 50.6 \\
RX0058 & B1 & 14.4012 & 26.7041 & 3.3 & 0.08 & -12. & -- & 28.9 \\
RX1740 & M & 265.1398 & 35.6416 & 2.8 & 0.21 & -26. & 0.17896 & 42.1 \\
RX1740 & S1 & 265.2600 & 35.4366 & 3.3 & 0.22 & 38. & 0.01946 & 31.7 \\
RX1740 & S2 & 264.8688 & 35.6053 & 3.7 & 0.52 & 28. & 0.01340 & 27.9 \\
RX1740 & S3 & 265.0744 & 35.8116 & 3.5 & 0.44 & -7. & 0.01166 & 21.8 \\
Z2844 & M & 150.7281 & 32.6483 & 2.9 & 0.20 & -85. & 0.10143 & 48.3 \\
Z2844 & S1 & 150.6524 & 32.7621 & 5.2 & 0.58 & 63. & 0.04930 & 50.5 \\
Z2844 & S2 & 150.5821 & 32.8890 & 2.6 & 0.39 & 0. & 0.00395 & 23.1 \\
Z8338 & M & 272.7447 & 49.9078 & 3.0 & 0.37 & -67. & 0.45876 & 43.1 \\
Z8338 & S1 & 272.8606 & 49.7916 & 3.2 & 0.11 & 62. & 0.05549 & 31.9 \\
Z8338 & S2 & 272.6903 & 49.9737 & 3.1 & 0.67 & 62. & 0.07089 & 25.7 \\
Z8338 & B1 & 272.4479 & 49.6815 & 1.7 & 0.18 & 9. & -- & 25.8 \\
Z8852 & M & 347.6024 & 7.5824 & 2.7 & 0.41 & -45. & 0.76110 & 67.9 \\
Z8852 & S1 & 347.5926 & 7.3999 & 5.8 & 0.56 & 62. & 0.12022 & 32.4 \\
Z8852 & S2 & 347.6986 & 7.8018 & 2.3 & 0.13 & 81. & 0.02493 & 25.5 \\
Z8852 & B1 & 347.7381 & 7.6808 & 2.1 & 0.25 & -73. & -- & 35.9 \\
Z8852 & B2 & 347.4951 & 7.8165 & 2.4 & 0.21 & 72. & -- & 27.0 \\
\hline
\label{tab:dedica}
\end{longtable}
\end{appendix}
\twocolumn

\end{document}